# Universal spectrum for atmospheric suspended particulates: comparison with observations

## A. M. Selvam[1]

**Deputy Director (Retired)**
**Indian Institute of Tropical Meteorology, Pune 411 008, India**
email: amselvam@gmail.com
Websites: http://amselvam.webs.com
http://amselvam.tripod.com/index.html

*Abstract*

Atmospheric flows exhibit self-similar fractal space-time fluctuations on all space-time scales in association with inverse power law distribution for power spectra of meteorological parameters such as wind, temperature, etc., and thus implies long-range correlations, identified as self-organized criticality generic to dynamical systems in nature. A general systems theory based on classical statistical physical concepts developed by the author visualizes the fractal fluctuations to result from the coexistence of eddy fluctuations in an eddy continuum, the larger scale eddies being the integrated mean of enclosed smaller scale eddies. The model satisfies the maximum entropy principle and predicts that the probability distributions of component eddy amplitudes and the corresponding variances (power spectra) are quantified by the same universal inverse power law distribution which is a function of the golden mean. Atmospheric particulates are held in suspension by the vertical velocity distribution (spectrum). The atmospheric particulate size spectrum is derived in terms of the model predicted universal inverse power law characterizing atmospheric eddy spectrum. Model predicted spectrum is in agreement with the following four experimentally determined data sets: (i) CIRPAS mission TARFOX_WALLOPS_SMPS aerosol size distributions (ii) CIRPAS mission ARM-IOP (Ponca City, OK) aerosol size distributions (iii) SAFARI 2000 CV-580 (CARG Aerosol and Cloud Data) cloud drop size distributions and (iv) TWP-ICE (Darwin, Australia) rain drop size distributions.

---

[1] Permanent address: Dr.Mrs.A.M.Selvam, B1 Aradhana, 42/2A Shivajinagar, Pune 411005, India. Tel: 09102025538194, Email: amselvam@gmail.com; websites: http://amselvam.webs.com; http://amselvam.tripod.com/index.html

## 1. **Introduction**

Atmospheric flows exhibit self-similar fractal space-time fluctuations on all space-time scales associated with inverse power law distribution or $1/\nu$ noise, where $\nu$ is the frequency, for power spectra of meteorological parameters such as wind, temperature, etc. Such $1/\nu$ noise imply long-range correlations, identified as self-organized criticality generic to dynamical systems in nature and are independent of the exact physical, chemical, physiological and other properties of the dynamical system. The physics of self-organized criticality is not yet identified.

Incidentally, the signature of fractals, namely, inverse power law form for power spectra of fluctuations was identified for isotropic homogeneous turbulence by Kolmogorov in the 1940s. The concept of fractals and its quantitative measure for space-time fluctuations of all scales was introduced by Mandelbrot in the late 1960s. The robust pattern of selfsimilar space-time fluctuations was identified by Bak, Tang and Wiesenfeld in the late 1980s as self-organised criticality (SOC) whereby the cooperative existence of fluctuations of all space-time scales maintains the dynamical equilibrium in a dynamical system. In this paper the author presents a general systems theory model applicable to all dynamical systems. The quantitative characteristics of the observed fractal space-time fluctuations and SOC are derived directly as a natural consequence of model concepts based on collective statistical probabilities of fluctuations such as in kinetic theory of gases as explained in the following. Visconti (2001) states that according to Edward Lorenz the atmosphere may be intrinsically unpredictable. Today there is no theory that could predict the evolution of a cloud in the presence of updraft, wind, humidity advection, etc. In a completely different context, the kinetic theory of gases solves another impossible problem and avoids the question of how to describe the exact position of each molecule in a gas. Instead it gives their collective properties, describing their statistical behavior (Visconti 2001). The most important problem of statistical mechanics is the kinetic theory of gases. Notions like pressure, temperature and entropy were based on the statistical properties of a large number of molecules (Contopoulos 2010). Recent work in dynamical systems theory has shown that many properties that are associated with irreversible processes in fluids can be understood in terms of the dynamical properties of reversible, Hamiltonian systems. That is, stochastic-like behavior is possible for these systems (Dorfman 1997). Maxwell's (1860s) and Boltzmann's (1870s) work on the kinetic theory of gases, and the creation of the more general theory of statistical mechanics persuaded many thinkers that certain very important large scale regularities – the various gas laws, and eventually the second law of thermodynamics – were indeed to be explained as the combined effect of the probability distributions governing those systems' parts (Strevens 2006).

A general systems theory developed by the author visualizes the fractal fluctuations to result from the coexistence of eddy fluctuations in an eddy continuum, the larger scale eddies being the integrated mean of enclosed smaller scale eddies. The model predicts that the probability distributions of component eddy amplitudes and the corresponding variances (power spectra) are quantified by the same universal inverse power law distribution which is a function of the golden mean. Atmospheric particulates are held in suspension by the vertical velocity distribution (spectrum). The atmospheric particulate size spectrum is derived in terms of the model predicted universal inverse power law characterizing atmospheric eddy energy spectrum.

Information on the size distribution of atmospheric suspended particulates (aerosols, cloud drops, raindrops) is important for the understanding of the physical processes relating to the studies in weather, climate, atmospheric electricity, air pollution and aerosol physics. Atmospheric suspended particulates affect the radiative balance of the Earth/atmosphere system via the direct effect whereby they scatter and absorb solar and terrestrial radiation, and via the indirect effect whereby they modify the microphysical properties of clouds thereby affecting the radiative properties and lifetime of clouds (Haywood et al. 2003). At present empirical models for the size distribution of atmospheric suspended particulates is used for quantitative estimation of earth-atmosphere radiation budget related to climate warming/cooling trends. The empirical models for different locations at different atmospheric conditions, however, exhibit similarity in shape implying a common universal physical mechanism governing the organization of the shape of the size spectrum. The pioneering studies during the last three decades by Lovejoy and his group (Lovejoy and Schertzer 2008, 2010) show that the particulates are held in suspension in turbulent atmospheric flows which exhibit selfsimilar fractal fluctuations on all scales ranging from turbulence (mm-sec) to climate (kms-years). Lovejoy and Schertzer (2008) have shown that the rain drop size distribution should show a universal scale invariant shape. In the present study a general systems theory for fractal space-time fluctuations developed by the author (Selvam 1990, 2005, 2007, 2009) is summarized and is applied to derive a universal (scale independent) spectrum for suspended atmospheric particulate size distribution expressed as a function of the golden mean $\tau$ ($\approx 1.618$), the total number concentration and the mean volume radius (or diameter) of the particulate size spectrum. A knowledge of the mean volume radius and total number concentration is sufficient to compute the total particulate size spectrum at any location. Model predicted spectrum is in agreement with the following four experimentally determined data sets: (i) CIRPAS mission TARFOX_WALLOPS_SMPS aerosol size distributions (ii) CIRPAS mission ARM-IOP (Ponca City, OK) aerosol size distributions (iii) SAFARI 2000 CV-580 (CARG Aerosol and Cloud Data) cloud drop size distributions and (iv) TWP-ICE (Darwin, Australia) rain drop size distributions. The paper is organized as follows. The current state of knowledge of the size distribution of atmospheric suspended particulates is given in Section 2 and Section 3 contains a brief summary of the observed characteristics of selfsimilar fractal fluctuations in atmospheric flows. Section 4 summarizes the general systems theory for fractal space-time fluctuations in atmospheric flows. The normalized (scale independent) atmospheric aerosol size spectrum is derived in Section 5. In Section 6 it is shown that the general systems theory concepts satisfy the maximum entropy principle of classical statistical physics. Sections 7 and 8 give respectively, details of observational data sets used for validating the theoretical predictions, and results with discussion of analyses of the data sets. The conclusions of the study are given in Section 9.

## 2. Atmospheric suspended particulates: current state of knowledge

### 2.1 Aerosol size distribution

As aerosol size is one of the most important parameters in describing aerosol properties and their interaction with the atmosphere, its determination and use is of fundamental importance. Aerosol size covers several decades in diameter and hence a variety of instruments are required for its determination. This necessitates several definitions of the diameter, the most common being the geometric diameter $d$. The size fraction with $d > 1$-$2$ μm is usually referred to as the coarse mode, and the fraction $d < 1$-$2$ μm is the fine mode. The latter mode can be further divided into the accumulation $d \sim 0.1$-$1$ μm, Aitken $d \sim 0.01$-$0.1$ um, and nucleation $d < 0.01$μm modes. Due to the $d^3$ dependence of aerosol volume (and mass), the coarse mode is typified by a maximum volume concentration and, similarly, the accumulation mode by the surface area concentration and the Aitken and nucleation modes by the number concentration.

Aerosol formation arises from heterogeneous or homogeneous nucleation. The former refers to condensation growth on existing nuclei, and the latter to the formation of new nuclei through condensation. Heterogeneous nucleation occurs preferentially on existing nuclei. Condensation onto a host surface occurs at a critical supersaturation, which is substantially lower (<1-2%) than for homogeneous nucleation in the absence of impurities (>300%). Examples of gas-to-particle conversion are combustion processes and the ambient formation of nuclei from gaseous organic emissions. Particles in the Aitken/accumulation mode typically arise from either: (i) the condensation of low volatility vapours; or (ii) coagulation. Particles in the accumulation mode have a longer atmospheric lifetime than other modes, as there is a minimum efficiency in sink processes. Particles in the coarse mode are usually produced by weathering and wind erosion processes. Dry deposition (primarily sedimentation) is the primary removal process. As the sources and sinks of the coarse and fine modes are different, there is only a weak association of particles in both modes (Hewitt and Jackson 2003). The aerosol chemistry data organized first by Peter Mueller and subsequently analyzed by Friedlander and coworkers showed that the fine and coarse mass modes were chemically distinctly different (Hussar 2005).

Husar (2005) has summarized the history of aerosol science as follows. The modern science of atmospheric aerosols began with the pioneering work of Christian Junge who performed the first comprehensive measurements of the size distribution and chemical composition of atmospheric aerosols (Junge 1952, 1953, 1955, 1963). Based on tedious and careful size distribution measurements performed over many different parts of the world, Junge and co-workers have observed that there is a remarkable similarity in the gathered size distributions (number concentration $N$ versus radius $r_a$): they follow a power law function over a wide range from 0.1 to over 20 μm in particle radius.

$$\frac{\mathrm{d}\,N}{\mathrm{d}\log r_a} = c r^{-\alpha}$$

The inverse power law exponent $\alpha$ of the number distribution function ranged between 3 and 5 with a typical value of 4. This power-law form of the size distribution became known as the ***Junge distribution*** of atmospheric aerosols. In the 1960s the physical mechanisms that were responsible for producing these similarities in the atmospheric aerosol size spectra were not known, although it was clear that homogeneous and heterogeneous nucleation, coagulation, sedimentation and other removal processes were all influential mechanisms. In particular, it was unclear which combination of these mechanisms is responsible for maintaining the observed ***quasi-stationary size distribution*** of the size spectra.

Whitby (1973) introduced the concept of the multimodal nature of atmospheric aerosol and Jaenicke and Davies (1976) added the mathematical formalism used today. Around 1970 - 71, Whitby et al. (1972) collected and analyzed several size distribution data sets arising from different locations, times, and sampling methods and the broad range of data provided strong evidence that bimodal distribution occurs as a ubiquitous feature of atmospheric aerosols in general, though the causal processes and mechanisms were unclear. Semi-quantitative explanation of the observed fine particle dynamics provided the scientific support for the bimodal concept and became the basis of regional dynamically coupled gas-aerosol models. As pointed out by Whitby (1978) and Junge (1963) an actual size distribution comes from the sum of single modes. There is an equivalency between the optical properties of a combination of several modes and a representative single mode. From previous work it can reasonably be assumed that aerosol size distributions follow a lognormal distribution (Tanre et al. 1996). Physical size distributions can be characterized well by a trimodal model consisting of three additive lognormal distributions (Whitey 2007). Typically, the planetary boundary layer (PBL) aerosol is combination of three modes corresponding to Aitken nuclei, accumulation mode aerosols, and coarse aerosols, the shape of which is often modeled as the sum of lognormal modes (Whitey 2007; Chen et al. 2009). In a nutshell, the bimodal distribution concept states that the atmospheric aerosol mass is distributed in two distinct size ranges, fine and coarse and that each aerosol mode has a characteristic size distribution, chemical composition and optical properties (Hussar 2005).

## 2.2 Cloud drop size distribution

### 2.2.1 Cloud microphysics and associated cloud dynamical processes

Prupaccher and Klett (1997) have summarized the current state of knowledge of cloud microphysical processes as follows. One principal continuing difficulty is that of incorporating, in a physically realistic manner, the microphysical phenomena in the broader context of the highly complex macrophysical environment of natural clouds. Mason (1957) also refers to the problem of scale in cloud microphysics. Cloud microphysics deals with the growth of particles ranging from the characteristic sizes of condensation nuclei ($\leq 10^{-2}$ µm) to precipitation particles ($\leq 10^4$ µm for raindrops, $\leq 10^5$ µm for hailstones). This means we must follow the evolution of the particle size spectrum, and the attendant microphysical processes of mass transfer, over about seven orders of magnitude in particle size. Similarly, the range of relevant cloud-air motions varies from the characteristic size of turbulent eddies which are small enough to decay directly through viscous dissipation ($\leq 10^{-2}$ cm), since it is these eddies which turn out to define the characteristic shearing rates for turbulent aerosol coagulation processes, to motion on scales at least as large as the cloud itself ($> 10^5$ cm). Thus, relevant interactions may occur over at least seven orders of magnitude of eddy sizes. A complete in-context understanding of cloud microphysics including dynamic, electrical and chemical effects is not yet available. Many microphysical mechanisms are still not understood in quantitative detail (Prupaccher and Klett 1997).

Although the relative humidity of clouds and fogs usually remains close to 100%, considerable departures from this value have been observed. The spatial and temporal non-uniformity of the humidity inside clouds and fogs results in a corresponding rapid spatial variation of the concentration of cloud drops and the cloud liquid water content. Based on his observations, Warner (1969) suggested that bimodal drop size distributions are the result of a mixing process between the cloud (cumulus) and the environment. Warner proposed that the mixing process producing the bimodality is due mostly to entrainment of drier air at the growing cloud top, and to a lesser degree, to entrainment at the cloud edges. The size distribution experiences a broadening effect with increasing distance from cloud base. Spectra with double maxima have also been observed by others in other regions. If we consider the spatial distribution of the drop size, number concentration, and liquid water content, we find strongly inhomogeneous conditions. The cloud liquid water content $w_L$ varies rapidly over short distances along a horizontal flight path in a manner which is closely related to the variation of the vertical velocity in the cloud and also $w_L$ varies essentially as the total number concentration of drops. Vulfson et al. (1973) demonstrate that the cloud water content typically increases with height above the cloud base, assumes a maximum somewhere in the upper half of the cloud, and then decreases again toward the cloud top. In most cases, a comparison between the observed cloud water content $w_L$ and adiabatic liquid water content $(w_L)_{ad}$ computed on the basis of a saturated adiabatic ascent of moist air shows that generally $w_L < (w_L)_{ad}$. In most cases, $w_L / (w_L)_{ad}$ is found to decrease with increasing height above cloud base but to increase with cloud width. This implies that the entrainment is especially pronounced near the cloud top, while the net dilution effect by entrainment is less in wider clouds than narrower ones.

### 2.2.2 Formulations for drop size distributions in clouds and fog

For many fog and cloud modeling purposes, it is necessary to be able to approximate the observed drop size distribution by an analytical expression. Fortunately, drop size distributions measured in many different types of clouds and fogs under a variety of meteorological conditions often exhibit a characteristic shape. Generally, the concentration rises sharply from a low value to a maximum, and then decreases gently towards larger sizes. Such a characteristic shape can be approximated reasonably well by either a gamma distribution or a lognormal distribution. In order to describe a drop size distribution with two or more maxima, one or more unimodal distributions may be superposed. As an example, according to Khrgian and Mazin (1952) (in Borovikov et al. 1963) many drop size distributions with a single maximum may also be quite well be represented by a gamma distribution. Another convenient representation of the cloud drop size distribution is the empirical formula developed by Best (1951a, b). These various analytical expressions only represent average distributions. Individual drop size spectra may be significantly different (Pruppacher and Klett 1997). A wealth of aircraft measurements in the Soviet Union indicate that droplet size spectra in stratocumulus are distributed in logarithmic normal (Levin 1954) or gamma forms (Borovikov et al. 1963). The droplet size spectra in stratus and stratocumulus are now commonly described by the gamma distribution. Droplet size spectra in altostratus and altocumulus as a function of temperature and cloud thickness are given according to Mazin and Khrgian (1989) (Hobbs 1993).

## 2.3 Rain drop size distribution

### 2.3.1 Classical cloud microphysical concepts

Rain drops are large enough to have a size dependent shape which cannot be characterized by a single length. The conventional resolution is to describe rain spectra in terms of the equivalent diameter $D_0$ defined as the diameter of a sphere of the same volume as the deformed drop. The overall shaping of the spectrum is obviously quite complicated, and determined in part by such meteorological variables as temperature, relative humidity, and wind in the subcloud region.

Various empirical relations have been advanced to describe the size spectra of raindrops. One often used is the size distribution proposed by Best (1950). Probably the most widely used description for the raindrop spectrum is the size distribution of Marshall and Palmer (MP) (1948), which is based on the observations of Laws and Parsons (1943). More detailed studies have demonstrated that the MP distribution is not sufficiently general to describe most observed raindrop spectra accurately. Numerous studies have also used the gamma distribution. Another alternative is the log normal distribution. Detailed comparison between the raindrop spectra actually observed and these empirical distributions show that in most cases only a partial fit can be achieved at best. The observed raindrop spectra also show, apart from a main mode, some secondary modes. It is reasonable to attribute the main mode as well as the subpeaks to collisional drop breakup. Unexpectedly, these peaks are not present in all raindrop size distributions. One explanation for this may be that the breakup-induced peaks become masked due to turbulent and evaporative effects. Additional factors which complicate an interpretation of observed raindrop distributions are related to instrumental problems (Pruppacher and Klett 1997).

### 2.3.2 Cloud microphysics and selfsimilar turbulent atmospheric flows

Lovejoy and his group (Lovejoy and Schertzer 2008, 2010) have done pioneering studies on selfsimilar fractal fluctuations ubiquitous to turbulent atmospheric flows and have emphasized the urgent need to incorporate, in modelling studies of microphysics of clouds and rain, the theory of nonlinear dynamical systems as summarized in the following. Rain is a highly turbulent process yet there is a wide gap between the turbulence and precipitation research. It is still common for turbulence to be invoked as a source of homogenization, an argument used to justify the use of homogeneous (white noise) Poisson process models of rain. Dimensional analysis shows that the cumulative probability distribution of non-dimensional drop mass should be a universal function dependent only on scale. Starting in the 1980s, a growing body of literature has demonstrated that—at least over large enough scales involving large numbers of drops—rain has nontrivial space–time scaling properties. While the traditional approach to drop modelling is to hypothesize specific parametric forms for the drop size distribution (DSD) and then to assume spatial homogeneity in the horizontal and smooth variations in the vertical, the nonlinear dynamics approach on the contrary assumes extreme turbulent-induced variability governed by the turbulent cascade processes and allows the DSD to be constrained by the turbulent fields. The conventional methods of modelling the evolution of raindrops give turbulence at most a minor (highly 'parameterized') role: the atmosphere is considered homogeneous and the spatial variability of the DSD arises primarily due to complex drop interactions. It is shown on dimensional grounds that the dimensionless cumulative DSD as a function of the dimensionless drop mass should be a universal function of dimensionless mass (Lovejoy and Schertzer 2008). Khain et al. (2007) have given critical comments to results of investigations of drop collisions in turbulent clouds and conclude that the fact that turbulence enhances the rate of particle collisions can be considered as being established.

## 3. Selfsimilar fractal fluctuations from turbulence to climate scales in atmospheric flows

The Atmospheric particulates are suspended in the selfsimilar wind fluctuation pattern ranging from turbulence to climate scales manifested as inverse power law form for power spectra of temporal fluctuations of wind speed. A brief summary of observed long-range correlations on all space-time scales in atmospheric flows and implications for modeling atmospheric dynamical transport processes is given in the following.

Atmospheric flows exhibit self-similar fractal fluctuations generic to dynamical systems in nature. Self-similarity implies long-range space-time correlations identified as self-organized criticality (Bak et al. 1988). The physics of self-organized criticality ubiquitous to dynamical systems in nature and in finite precision computer realizations of non-linear numerical models of dynamical systems is not yet identified. During the past three decades, Lovejoy and his group (Lovejoy and Schertzer 2010) have done extensive observational and theoretical studies of fractal nature of atmospheric flows and emphasized the urgent need to formulate and incorporate quantitative theoretical concepts of fractals in mainstream classical theory relating to Atmospheric Physics.

The empirical analyses summarized by Lovejoy and Schertzer (2010), Bunde et al. (2003), Bunde and Havlin (2002, 2003), Eichner et al. (2003), Rybski et al. (2006), Rybski, et al. (2008), directly demonstrate the strong scale dependencies of many atmospheric fields, showing that they depend in a power law manner on the space–time scales over which they are measured. In spite of intense efforts over more than 50 years, analytic approaches have been surprisingly ineffective at deducing the statistical properties of turbulence. Atmospheric Science labors under the misapprehension that its basic science issues have long been settled and that its task is limited to the application of known laws — albeit helped by ever larger quantities of data themselves processed in evermore powerful computers and exploiting ever more sophisticated algorithms. Conclusions about anthropogenic influences on the atmosphere can only be drawn with respect to the null hypothesis, i.e. one requires a theory of the natural variability, including knowledge of the probabilities of the extremes at various resolutions. At present, the null hypotheses are classical so that they assume there are no long range statistical dependencies and that the probabilities are thin-tailed (i.e. exponential). However observations show that cascades involve long-range dependencies and (typically) have fat tailed (algebraic) distributions in which extreme events occur much more frequently and can persist for much longer than classical theory would allow (Lovejoy and Schertzer 2010; Bogachev, Eichner, and Bunde 2008a, b; Eichner, Kantelhardt, Bunde, and Havlin 2006; Bunde, Eichner, Kantelhardt, and Havlin 2005; Bogachev, Eichner, and Bunde 2007).

A general systems theory for the observed fractal space-time fluctuations of dynamical systems developed by the author (Selvam 1990, 2007) helps formulate a simple model to explain the observed vertical distribution of number concentration and size spectra of atmospheric aerosols. The atmospheric aerosol size spectrum is derived in terms of the universal inverse power law characterizing atmospheric eddy energy spectrum. The physical basis and the theory relating to the model are discussed in Section 4. The model predictions are (i) The fractal fluctuations can be resolved into an overall logarithmic spiral trajectory with the quasiperiodic Penrose tiling pattern for the internal structure. (ii) The probability distribution of fractal space-time fluctuations (amplitude) also represents the power (variance or square of amplitude) spectrum for fractal fluctuations and is quantified as universal inverse power law incorporating the *golden mean*. Such a result that the additive amplitudes of eddies when squared represent probability distribution is observed in the subatomic dynamics of quantum systems such as the electron or photon. Therefore the irregular or unpredictable fractal fluctuations exhibit quantum-like chaos. (iii) Atmospheric aerosols are held in suspension by the vertical velocity fluctuation distribution (spectrum). The normalized (scale independent) atmospheric aerosol size spectrum is derived in terms of the universal inverse power law characterizing atmospheric eddy energy spectrum. Model predicted spectrum is in agreement (within two standard deviations on either side of the mean) with experimentally determined data sets (Sections 7 and 8)

## 4. General systems theory for fractal space-time fluctuations in atmospheric flows

The study of the spontaneous, i.e., self-organized formation of structures in systems far from thermal equilibrium in open systems belongs to the multidisciplinary field of *synergetics* (Haken, 1989). Formation of structure begins by aggregation of molecules in a turbulent fluid (gas or liquid) medium. Turbulent fluctuations are therefore not dissipative, but serve to assemble and form coherent structures (Nicolis and Prigogine 1977; Prigogine 1980; Prigogine and Stengers 1988; Insinnia 1992), for example, the formation of clouds in turbulent atmospheric flows. Traditionally, turbulence is considered dissipative and disorganized. Yet, coherent (organized) vortex roll circulations (vortices) are ubiquitous to turbulent fluid flows (Levich 1987; Frisch and Orszag 1990). The exact physical mechanism for the formation and maintenance of coherent structures, namely vortices or large eddy circulations in turbulent fluid flows is not yet identified.

Turbulence, namely, seemingly random fluctuations of all scales, therefore, plays a key role in the formation of selfsimilar coherent structures in the atmosphere. Such a concept is contrary to the traditional view that turbulence is dissipative, i.e., ordered growth of coherent form is not possible in turbulent flows. The author (Selvam, 1990, 2007) has shown that turbulent fluctuations self-organize to form selfsimilar structures in fluid flows.

In summary, spatial integration of enclosed turbulent fluctuations give rise to large eddy circulations in fluid flows. Therefore, starting with turbulence scale fluctuations, progressively larger scale eddy fluctuations can be generated by integrating circulation structures at different scale ranges. Such a concept envisages only the magnitude (intensity) of the fluctuations and is independent of the properties of the medium in which the fluctuations are generated. Also, self-similar space-time growth structure is implicit to hierarchical growth process, i.e., the large scale structure is the envelope of enclosed smaller scale structures. Successively larger scale structures form a hierarchical network and function as a unified whole.

The role of surface frictional turbulence in weather systems is discussed in Section 4.1. The common place occurrence of long-lived organized cloud patterns and their important contribution to the radiation budget of the earth's atmosphere is briefly discussed in Section 4.1.1.

## 4.1 Frictional convergence induced weather

Roeloffzen et al. (1986) discussed the importance of frictional convergence induced weather as follows. The coastline generally represents a marked discontinuity in surface roughness. The resulting mechanical forcing leads to a secondary circulation in the boundary layer, and consequently to a vertical motion field that may have a strong influence on the weather in the coastal zone. In potentially unstable air masses, frictional convergence may cause a more-or-less stationary zone of heavy shower activity, for example. Of all meteorological phenomena typical for coastal regions, the fair weather sea-breeze circulation has probably been studied most extensively, e.g., Estoque (1962), Walsh (1974), Pielke (1974), Pearson et al. (1983). In contrast to this thermally-driven circulation, the mechanical forcing due to the discontinuity in surface roughness may create circulation patterns of similar amplitude and scale. Frictional convergence is mentioned in some studies as the cause of increased precipitation in coastal zones under specific conditions, e.g., Bergeron (1949), Timmerman (1963), Oerlemans (1980); but its effect is generally underestimated (Roeloffzen et al. 1986).

Frictional convergence is analogous to Ekman pumping, namely, the process of inducing vertical motions by boundary layer friction (Stull 1988).

Cotton and Anthes (1989) emphasized the importance of the role of Ekman pumping on large scale weather systems. The strong control of cumulus convection by the larger scales of motion in tropical cyclones has been recognized for a long time. Syono et al. (1951) showed that the rate of precipitation in typhoons was related to the updrafts produced by frictional convergence in the PBL (so-called Ekman pumping). Later observational and modeling studies have confirmed the cooperative interaction between cumulus convection and the tropical cyclones through frictionally induced moisture convergence and enhanced evaporation in the PBL (see review in Anthes 1982). Local winds such as Sea breezes actually are very important because they are determined mainly by the interaction of large scale motions with local topography (Visconti 2001).

New research suggests that rough areas of land, including city buildings and naturally jagged land cover like trees and forests can actually attract passing hurricanes. It was observed that storms traveling over river deltas hold together longer than those over dry ground. As a result, the city of New Orleans might feel a greater impact of hurricanes coming off the Gulf of Mexico than existing computer models predict (Au-Yeung et al., 2010; Lucibella 2010)

### 4.1.1 Mesoscale cellular convection and radiation budget of the earth

Feingold et al. (2010) discussed the importance of the observed large scale organized pattern of clouds in the radiation budget of the earth's atmosphere. Cloud fields adopt many different patterns that can have a profound effect on the amount of sunlight reflected back to space, with important implications for the Earth's climate. These cloud patterns can be observed in satellite images of the Earth and often exhibit distinct cell-like structures associated with organized convection at scales of tens of kilometers (Krueger and Fritz 1961; Agee 1984; Garay et al. 2004), i.e. mesoscale cellular convection. These clouds are important because they increase the reflectance of shortwave radiation and therefore exert a cooling effect on the climate system that is not compensated by appreciable changes in outgoing longwave radiation (Twomey 1977).

## 4.2 Growth of macro-scale coherent structures from microscopic domain fluctuations in atmospheric flows

The non-deterministic model (Selvam 1990, 2007, 2009) described below incorporates the physics of the growth of macro-scale coherent structures from microscopic domain fluctuations in atmospheric flows. In summary, the mean flow at the planetary ABL possesses an inherent upward momentum flux of frictional origin at the planetary surface. This turbulence-scale upward momentum flux is progressively amplified by the exponential decrease of the atmospheric density with height coupled with the buoyant energy supply by micro-scale fractional condensation on hygroscopic nuclei, even in an unsaturated environment (Pruppacher and Klett 1997). The mean large-scale upward momentum flux generates helical vortex-roll (or large eddy) circulations in the planetary atmospheric boundary layer and under favourable conditions of moisture supply, is manifested as cloud rows and (or) streets, and mesoscale cloud clusters MCC in the global cloud cover pattern. A conceptual model of large and turbulent eddies in the planetary ABL is shown in Figs. 1 and 2. The mean airflow at the planetary surface carries the signature of the fine scale features of the planetary surface topography as turbulent fluctuations with a net upward momentum flux. This persistent upward momentum flux of surface frictional origin generates large-eddy (or vortex-roll) circulations, which carry upward the turbulent eddies as internal circulations. Progressive upward growth of a large eddy occurs because of buoyant energy generation in turbulent fluctuations as a result of the latent heat of condensation of atmospheric water vapour on suspended hygroscopic nuclei such as common salt particles. The latent heat of condensation generated by the turbulent eddies forms a distinct warm envelope or a micro-scale capping inversion layer at the crest of the large-eddy circulations as shown in Figure 1.

Progressive upward growth of the large eddy occurs from the turbulence scale at the planetary surface to a height $R$ and is seen as the rising inversion of the daytime atmospheric boundary layer (Figure 2).

The turbulent fluctuations at the crest of the growing large-eddy mix overlying environmental air into the large-eddy volume, i.e. there is a two-stream flow of warm air upward and cold air downward analogous to superfluid turbulence in liquid helium (Donnelly 1988, 1990). The convective growth of a large eddy in the atmospheric boundary layer therefore occurs by vigorous counter flow of air in turbulent fluctuations, which releases stored buoyant energy in the medium of propagation, e.g. latent heat of condensation of atmospheric water vapour. Such a picture of atmospheric convection is different from the traditional concept of atmospheric eddy growth by diffusion, i.e. analogous to the molecular level momentum transfer by collision.

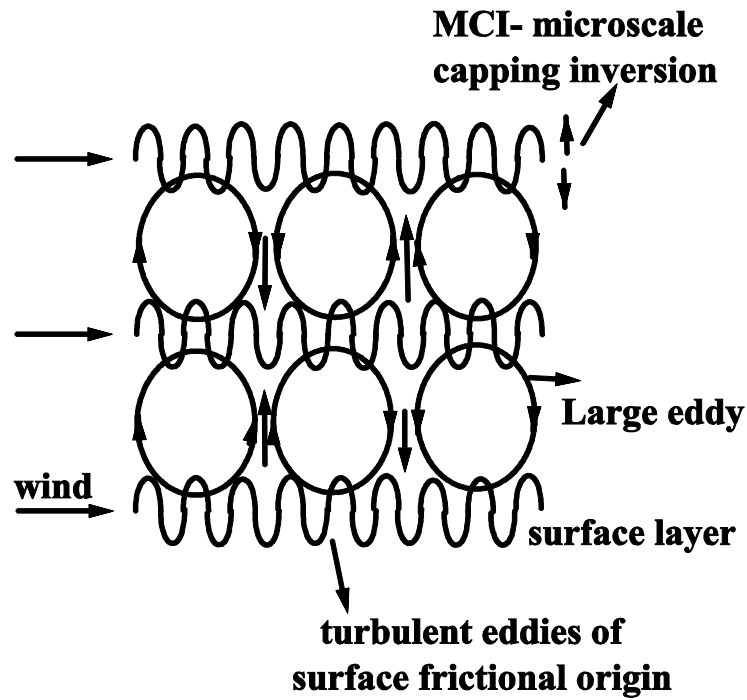

Fig. 1. Micro-scale capping inversion (MCI) layer at the crest of the large-eddy circulations

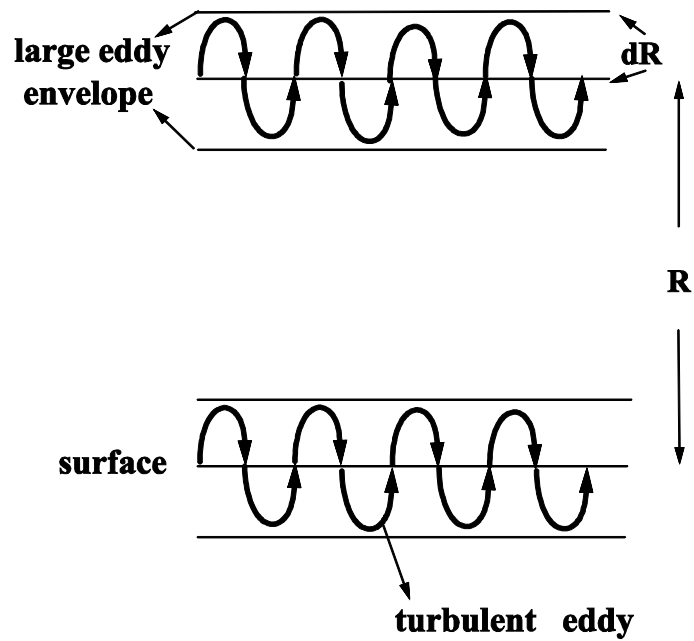

Fig. 2. Progressive upward growth of the large eddy from the turbulence scale at the planetary surface.

The generation of turbulent buoyant energy by the micro-scale fractional condensation is maximum at the crest of the large eddies and results in the warming of the large-eddy volume. The turbulent eddies at the crest of the large eddies are identifiable by a micro-scale capping inversion that rises upward with the convective growth of the large eddy during the course of the day. This is seen as the rising inversion of the daytime planetary boundary layer in echosonde and radiosonde records and has been identified as the entrainment zone (Boers 1989; Gryning and Batchvarova 2006) where mixing with the environment occurs.

The general systems theory for eddy growth discussed so far for planetary atmospheric boundary layer (ABL) can be extended up to the upper atmospheric levels. In summary, a gravity wave feedback mechanism for the vertical mass exchange between the troposphere and the stratosphere is proposed. The vertical mass exchange takes place through a chain of eddy systems. The atmospheric boundary layer (ABL) contains large eddies (vortex rolls) which carry on their envelopes turbulent eddies of surface frictional origin (Selvam et al. 1984a; Selvam 1990, 2007). The buoyant energy production by *microscale-fractional-condensation* (MFC) in turbulent eddies is responsible for the sustenance and growth of large eddies (Selvam et al. 1984b; Selvam 1990, 2007). The buoyant energy production of turbulent eddies by the *microscale-fractional-condensation* (MFC) process is maximum at the crest of the large eddies and results in the warming of the large eddy volume. The turbulent eddies at the crest of the large eddies are identifiable by a *microscale-capping-inversion* (MCI) layer which rises upwards with the convective growth of the large eddy in the course of the day. The MCI layer is a region of enhanced aerosol concentrations. As the *microscale-fractional-condensation* (MFC) generated warm parcel of air corresponding to the large eddy rises in the stable environment of the *microscale-capping-inversion* (MCI), *Brunt Vaisala* oscillations are generated (Selvam et al., 1984b; Selvam, 1990, 2007). The growth of the large eddy is associated with generation of a continuous spectrum of gravity (buoyancy) waves in the atmosphere. The atmosphere contains a stack of large eddies. Vertical mixing of overlying environmental air into the large eddy volume occurs by turbulent eddy fluctuations (Selvam et al. 1984a; Selvam 1990, 2007). The circulation speed of the large eddy is related to that of the turbulent eddy according to the following expression (Townsend 1956; Selvam 1990).

$$W^2 = \frac{2}{\pi} \frac{r}{R} w^2 \qquad (1)$$

In the above Eq. 1 $W$ and $w$ are respectively the r.m.s (root mean square) circulation speeds of the large and turbulent eddies and $R$ and $r$ are their respective radii.

The relationship between the time scales $T$ and $t$ respectively of the large and turbulent eddies can be derived in terms of the circulation speeds $W$ and $w$ and their respective length scales $R$ and $r$ from Eq. 1 (Selvam 1990) as follows.

$$T = \frac{2\pi R}{W} \quad \text{and} \quad t = \frac{2\pi r}{w}$$

$$\frac{T}{t} = \frac{R}{r} \frac{w}{W} = \frac{R}{r} \sqrt{\frac{\pi}{2} \frac{R}{r}} = \left(\frac{R}{r}\right)^{\frac{3}{2}} \sqrt{\frac{\pi}{2}}$$

As seen from Figs. 1 and 2 and from the concept of eddy growth, vigorous counter flow (mixing) characterizes the large-eddy volume. The total fractional volume dilution rate of the large eddy by vertical mixing across unit cross-section is derived from Eq. 1 (Selvam et al. 1984a; Selvam 1990, 2007) and is given as follows.

$$k = \frac{w}{dW} \frac{r}{R} \qquad (2)$$

In Eq. 2 $w$ is the increase in vertical velocity per second of the turbulent eddy due to microscale fractional condensation (MFC) process and $dW$ is the corresponding increase in vertical velocity of large eddy.

The fractional volume dilution rate $k$ is equal to 0.4 for the scale ratio ($z$) $R/r =10$. Identifiable large eddies can exist in the atmosphere for scale ratios more than 10 only since, for smaller scale ratios the fractional volume dilution rate $k$ becomes more than half. Thus atmospheric eddies of various scales, i.e., convective, meso-, synoptic and planetary scale eddies are generated by successive decadic scale range eddy mixing process starting from the basic turbulence scale (Selvam et al. 1984b; Selvam 1990, 2007).

From Eq. 2 the following logarithmic wind profile relationship for the *ABL* is obtained (Selvam et al. 1984a; Selvam 1990, 2007).

$$W = \frac{w}{k} \ln z \qquad (3)$$

The steady state fractional upward mass flux $f$ of surface air at any height $z$ can be derived using Eq. 3 and is given by the following expression (Selvam et al. 1984a; Selvam 1990, 2007).

$$f = \sqrt{\frac{2}{\pi z}} \ln z \qquad (4)$$

In Eq. 4 $f$ represents the steady state fractional volume of surface air at any level $z$. Since atmospheric aerosols originate from surface, the vertical profile of mass and number concentration of aerosols follow the $f$ distribution.

The magnitude of the steady state vertical aerosol mass flux is dependent on $m_*$, the aerosol mass concentration at the initial level (earth's surface) and is equal to $m_* f$ from Eq. 4, the non-zero values of $f$ being given in terms of the non-dimensional length scale ratio $z$.

vertical distribution of atmospheric aerosol concentration

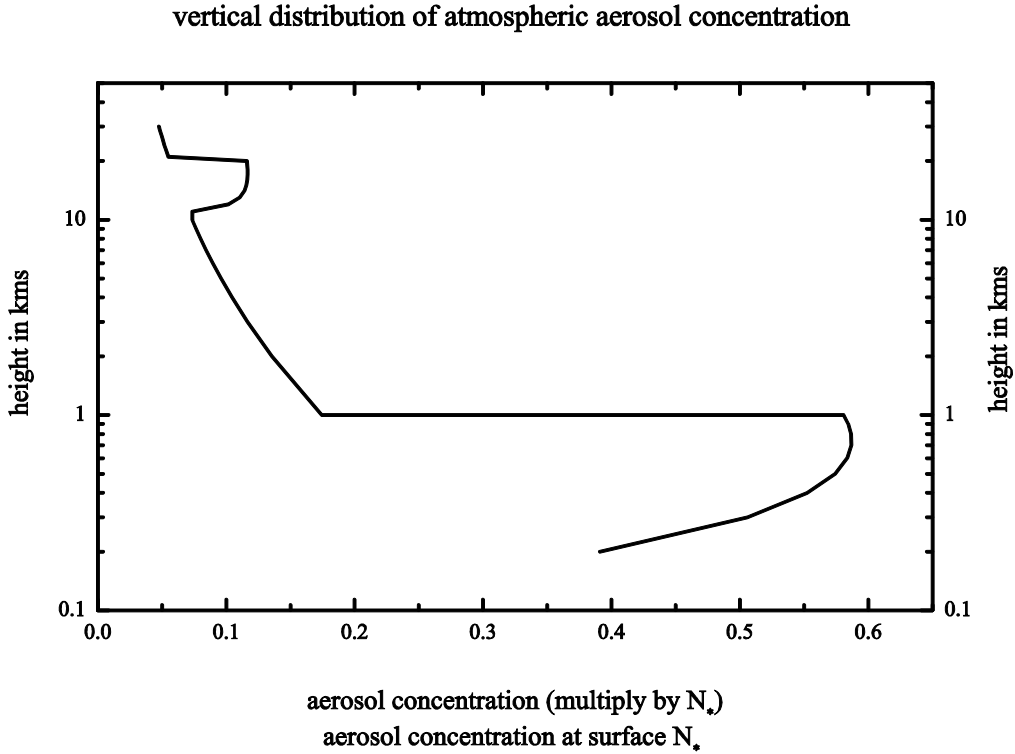

Fig. 3. Model predicted aerosol vertical distribution

The aerosol concentration vertical profile at Figure 3 is computed using Eq. 4 with appropriate length scale ratio $z$ values corresponding to the associated steady state fractional volume dilution $k$ values (Eq. 2). The fractional volume dilution rate $k$ is equal to 0.4 for the scale ratio ($z$) $R/r$ =10. Identifiable large eddies can exist in the atmosphere only for scale ratios more than 10 since, for smaller scale ratios the fractional volume dilution rate $k$ becomes more than half. Thus atmospheric eddies of various scales, i.e., convective, meso-, synoptic and planetary scale eddies are generated by successive decadic scale range eddy mixing process starting from the basic turbulence scale (Selvam et al. 1984a, b; 1992; 1996; Selvam and Joshi 1995 Selvam and Fadnavis 1998; Joshi and Selvam 1999; Selvam 1990, 1993, 2005, 2007, 2009, 2011).

The peaks in the aerosol concentration at 1 km (*lifting condensation level*) and at about 10-15 km (*stratosphere*) identify the microscale capping inversion (MCI, Figure 1) at the crests of the convective and meso-scale eddies respectively, the appropriate scale ratios for the convective and meso-scale eddies being 10 and 100 with respect to the turbulence scale. Thus for the turbulent eddy of radius 100m, the MCI's for the convective and meso-scale eddies occur at 1 km and 10 km respectively.

The model predicted profiles closely resemble the observed profiles associated with quasi-permanent tropospheric inversion (temperature) layers reported by other investigators (Junge 1963).

The vertical mass exchange mechanism predicts the $f$ distribution for the steady state vertical transport of aerosols at higher levels. Thus aerosol injection into the stratosphere by volcanic eruptions gives rise to the enhanced peaks in the regions of microscale capping inversion (MCI) in the stratosphere and other higher levels determined by the radius of the dominant turbulent eddy at that level.

The time $T$ taken for the steady state aerosol concentration $f$ to be established at the normalised height $z$ is equal to the time taken for the large eddy to grow to the height $z$ and is computed using the following relation (see Section 4.3.2 below).

$$T = \frac{r}{w} \sqrt{\frac{\pi}{2}} \, li \sqrt{z} \qquad (5)$$

In Eq. 5 $li$ is the logarithm integral.

The vertical dispersion rate of aerosols/pollutants from known sources (e.g., volcanic eruptions, industrial emissions) can be computed using the relation for $f$ and $T$ (Eqs. 4 and 5).

## 4.3 Computations of model predictions and comparison with observations

### 4.3.1 Vertical velocity profile

The microscale fractional condensation (MFC) generated values of vertical velocity have been calculated for different heights above the surface for clear-air conditions and above the cloud-base for in-cloud conditions for a representative tropical environment with favourable moisture supply. A representative cloud-base height is considered to be 1000m above sea level (a.s.l) and the corresponding meteorological parameters are, surface pressure 1000 mb, surface temperature 30$^{o}$C, relative humidity at the surface 80%, turbulent length scale 1 cm. The values of the latent heat of vapourisation $L_V$ and the specific heat of air at constant pressure $C_p$ are 600 cal gm$^{-1}$ and 0.24 cal gm$^{-1}$ respectively. The ratio values of $m_w/m_0$, where $m_0$ is the mass of the hygroscopic nuclei per unit volume of air and $m_w$ is the mass of water condensed on $m_0$, at various relative humidities as given by Winkler and Junge (1971, 1972) have been adopted and the value of $m_w/m_0$ is equal to about 3 for relative humidity 80%. For a representative value of $m_0$ equal to 100µg m$^{-3}$ the temperature perturbation $\theta'$ is equal to 0.00065$^{o}$C and the corresponding vertical velocity perturbation (turbulent) $w*$ is computed and is equal to 21.1x10$^{-4}$ cm sec$^{-1}$ from the following relationship between the corresponding virtual potential temperature $\theta_v$, and the acceleration due to gravity g equal to 980.6 cm sec$^{-2}$.

$$w_* = \frac{g}{\theta_v} \theta'$$

Heat generated by condensation of water equal to 300 µg on 100 µg of hygroscopic nuclei per meter[3] generates vertical velocity perturbation $w_*$ (cm sec$^{-2}$) equal to $21.1 \times 10^{-4}$ cm sec$^{-2}$ at surface levels. In the following it is shown that a value of $w_*$ equal to $30 \times 10^{-7}$ cm sec$^{-2}$, i.e. about three orders of magnitude less than that shown in the above example is sufficient to generate clouds as observed in practice.

From the logarithmic wind profile relationship (Eq. 3) and the steady state fractional upward mass flux $f$ of surface air at any height $z$ (Eq. 4) the corresponding vertical velocity perturbation $W$ can be expressed in terms of the primary vertical velocity perturbation $w_*$ as

$$W = w_* f z \qquad (6)$$

$W$ may be expressed in terms of the scale ratio $z$ as given below

From Eq. 4
$$f = \sqrt{\frac{2}{\pi z}} \ln z$$

Therefore
$$W = w_* z \sqrt{\frac{2}{\pi z}} \ln z = w_* \sqrt{\frac{2z}{\pi}} \ln z$$

The values of large eddy vertical velocity perturbation $W$ produced by the process of microscale fractional condensation at normalized height $z$ computed from Eq. 6 are given in the Table 1.

| Table 1 | |
|---|---|
| Height above surface $z$ | Vertical velocity $W$ cm sec$^{-1}$ |
| 1 cm | $30 \times 10^{-7}$ |
| 100 cm | $1.10 \times 10^{-4}$ |
| 100 m | $2.20 \times 10^{-3}$ |
| 1000 m | $8.71 \times 10^{-3} \approx 0.01$ |

The above values of vertical velocity although small in magnitude are present for long enough period in the lower levels and contribute for the formation and development of clouds as explained below.

### 4.3.2 Large eddy growth time

The time required for the large eddy of radius $R$ to grow from the primary turbulence scale radius $r_*$ is computed as follows.

The scale ratio $z = \dfrac{R}{r_*}$

Therefore for constant turbulence radius $r_*$

$$\mathrm{d}\,z = \frac{\mathrm{d}\,R}{r_*}$$

The incremental growth d$R$ of large eddy radius is equal to

$$\mathrm{d}\,R = r_* \,\mathrm{d}\,z$$

The time d$t$ for the incremental cloud growth is expressed as follows

$$\mathrm{d}\,t = \frac{\mathrm{d}\,R}{W} = \frac{r_* \,\mathrm{d}\,z}{W}$$

$W$ is the increase in large eddy circulation speed resulting from enclosed turbulent eddy circulations of speed $w_*$ and is given as $W = w_* f z$ from Eq. 6. Therefore

$$\mathrm{d}\,t = \frac{r_* \,\mathrm{d}\,z}{w_* f z} = \frac{r_* \,\mathrm{d}\,z}{w_* z \sqrt{\dfrac{2}{\pi z}} \ln z}$$

$$t = \frac{r_*}{w_*} \sqrt{\frac{\pi}{2}} \int\limits_{2}^{z} \frac{\mathrm{d}\,z}{z^{1/2} \ln z}$$

The above equation can be written in terms of $\sqrt{z}$ as follows

$$\mathrm{d}(z^{0.5}) = \frac{\mathrm{d}\,z}{2\sqrt{z}}$$

$$\mathrm{d}\,z = 2\sqrt{z}\,\mathrm{d}(\sqrt{z})$$

Therefore

$$t = \frac{r_*}{w_*} \sqrt{\frac{\pi}{2}} \int\limits_{x1}^{x2} \frac{\mathrm{d}(\sqrt{z})}{\ln \sqrt{z}} = \frac{r_*}{w_*} \sqrt{\frac{\pi}{2}} \int\limits_{x1}^{x2} \mathrm{li}\!\left(\sqrt{z}\right) \qquad (7)$$

$$x_1 = \sqrt{z_1} \ \text{ and } \ x_2 = \sqrt{z_2}$$

In the above equation $z_1$ and $z_2$ refer respectively to lower and upper limits of integration and li is the Soldner's integral or the logarithm integral. The large eddy growth time $t$ can be computed from Eq. 7.

The time $t$ seconds taken for the evolution of the 1000m ($10^5$ cm) eddy from the 1cm radius ($r_*$) eddy at the surface energized by the *microscale fractional condensation* (MFC) induced primary perturbation $w_*$ equal to 0.01cm sec$^{-2}$ at the surface levels can be computed from the above equation by substituting for $z_1 = 1$cm and $z_2 = 10^5$ cm such that $x_1 = \sqrt{1} = 1$ and $x_2 = \sqrt{10^5} \approx 317$.

$$t = \frac{1}{0.01} \sqrt{\frac{\pi}{2}} \int\limits_{1}^{317} \mathrm{li}(z)$$

The value of $\int\limits_{1}^{317} \mathrm{li}(z)$ is equal to 71.3

Hence $t \approx 8938$ sec $\approx 2$ hrs 30 mins

Thus starting from the surface level cloud growth begins after a time period of 2 hrs 30 mins. This is consistent with the observations.

# 5. Atmospheric aerosol size spectrum

## 5.1 Vertical variation of aerosol number concentration

The atmospheric eddies hold in suspension the aerosols and thus the mass size spectrum of the atmospheric aerosols is dependent on the vertical velocity fluctuation spectrum of the atmospheric eddies as explained in the following. The distribution of atmospheric aerosols in not only determined by turbulence, but also by dry and wet chemistry, sedimentation, gas to particle conversion, coagulation, (fractal) variability at the surface, amongst others. However, at any instant, the mass (and therefore the radius for homogeneous aerosols) size distribution of atmospheric suspensions (aerosols) is directly related to the wind vertical velocity (eddy energy) spectrum which is shown to be universal (scale independent). The source for aerosols in the fine mode (less than 1 $\mu$m) and coarse mode (greater than 1 $\mu$m) are different and may account for the differences in the observed departures of the observed from model predicted radius size spectrum for the fine and coarse aerosol modes (Section 7).

From the logarithmic wind profile relationship (Eq. 3) and the steady state fractional upward mass flux $f$ of surface air at any height $z$ (Eq. 4) the vertical velocity $W$ is expressed in Eq. 6 as

$$W = w_* f z$$

The corresponding moisture content $q$ at height $z$ is related to the moisture content $q_*$ at the surface (or reference level) and is given as (from Eq. 6)

$$q = q_* f z \qquad (7)$$

The aerosols are held in suspension by the eddy vertical velocity perturbations. Thus the suspended aerosol mass concentration $m$ at any level $z$ will be directly related to the vertical velocity perturbation $W$ at $z$, i.e., $W \sim mg$ where $g$ is the acceleration due to gravity. Therefore

$$m = m_* f z \qquad (8)$$

In Eq. 8 $m_*$ is the suspended aerosol (homogeneous) mass concentration in the surface layer. Let $r_a$ and $N$ represent the mean volume radius and number concentration of aerosols at level $z$. The variables $r_{as}$ and $N_*$ relate to corresponding parameters at the surface levels. Substituting for the average mass concentration in terms of mean radius $r_a$ and number concentration $N$ at normalized height $z$ above surface

$$\frac{4}{3}\pi r_a^3 N = \frac{4}{3}\pi r_{as}^3 N_* f z \qquad (9)$$

The number concentration $N$ of aerosol decreases with normalised height $z$ according to the $f$ distribution as shown earlier in Section 4.2 and is expressed as follows:

$$N = N_* f \qquad (10)$$

## 5.2 Vertical variation of aerosol mean volume radius

The mean volume radius of aerosol increases with height (eddy radius) $z$ as shown in the following. At any height $z$, the fractal fluctuations (of wind, temperature, etc.) carry the signatures of eddy fluctuations of all size scales since the eddy of length scale $z$ encloses smaller scale eddies and at the same time forms part of internal circulations of eddies larger than length scale $z$.

The wind velocity perturbation $W$ is represented by an eddy continuum of corresponding size (length) scales $z$. The aerosol mass flux across unit cross-section per unit time is obtained by normalizing the velocity perturbation $W$ with respect to the corresponding length scale $z$ to give the volume flux of air equal to $Wz$ and can be expressed as follows from Eq. 6:

$$Wz = \left(w_* fz\right)z = w_* fz^2 \tag{11}$$

The corresponding normalized moisture flux perturbation is equal to $qz$ where $q$ is the moisture content per unit volume at level $z$. Substituting for $q$ from Eq. 7

$$\text{normalised moisture flux at level } z = q_* fz^2 \tag{12}$$

The moisture flux increases with height resulting in increase of mean volume radius of CCN because of condensation of water vapour. The corresponding CCN (aerosol) mean volume radius $r_a$ at height $z$ is given in terms of the aerosol number concentration $N$ at level $z$ and mean volume radius $r_{as}$ at the surface (or reference level) as follows from Eq. 12

$$\frac{4}{3}\pi r_a^3 N = \frac{4}{3}\pi r_{as}^3 N_* fz^2 \tag{13}$$

Substituting for $N$ from Eq. 10 in terms of $N_*$ and $f$

$$r_a^3 = r_{as}^3 z^2$$
$$r_a = r_{as} z^{2/3} \tag{14}$$

The mean aerosol size increases with height according to the cube root of $z^2$ (Eq. 14). As the large eddy grows in the vertical, the aerosol size spectrum extends towards larger sizes while the total number concentration decreases with height according to the $f$ distribution. The atmospheric aerosol size spectrum is dependent on the eddy energy spectrum and may be expressed in terms of the recently identified universal characteristics of fractal fluctuations generic to atmospheric flows (Selvam 2009, 2010) as shown in Section 5.3 below.

## 5.3 Probability distribution of fractal fluctuations in atmospheric flows

The atmospheric eddies hold in suspension the aerosols and thus the size spectrum of the atmospheric aerosols is dependent on the vertical velocity spectrum of the atmospheric eddies. Atmospheric air flow is turbulent, i.e., consists of irregular fluctuations of all space-time scales characterized by a broadband spectrum of eddies. The suspended aerosols will also exhibit a broadband size spectrum closely related to the atmospheric eddy energy spectrum.

Atmospheric flows exhibit self-similar fractal fluctuations generic to dynamical systems in nature such as fluid flows, heart beat patterns, population dynamics, spread of forest fires, etc. Power spectra of fractal fluctuations exhibit inverse power law of form $v^{-\alpha}$ where $\alpha$ is a constant indicating long-range space-time correlations or persistence. Inverse power law for power spectrum indicates scale invariance, i.e., the eddy energies at two different scales (space-time) are related to each other by a scale factor ($\alpha$ in this case) alone independent of the intrinsic properties such as physical, chemical, electrical etc of the dynamical system.

A general systems theory for turbulent fluid flows predicts that the eddy energy spectrum, i.e., the variance (square of eddy amplitude) spectrum is the same as the probability distribution $P$ of the eddy amplitudes, i.e. the vertical velocity $W$ values. Such a result that the additive amplitudes of eddies, when squared, represent the probabilities is exhibited by the subatomic dynamics of quantum systems such as the electron or photon. Therefore the unpredictable or irregular fractal space-time fluctuations generic to dynamical systems in nature, such as atmospheric flows is a signature of quantum-like chaos. The general systems theory for turbulent fluid flows predicts (Selvam 1990, 2005, 2007) that the atmospheric eddy energy spectrum follows inverse power law form incorporating the *golden mean* $\tau$ (Selvam 2009) and the normalized deviation $\sigma$ for values of $\sigma \geq 1$ and $\sigma \leq -1$ as given below

$$P = \tau^{-4\sigma} \tag{15}$$

The vertical velocity $W$ spectrum will therefore be represented by the probability distribution $P$ for values of $\sigma \geq 1$ and $\sigma \leq -1$ given in Eq. 15 since fractal fluctuations exhibit quantum-like chaos as explained above.

$$W = P = \tau^{-4\sigma} \tag{16}$$

Values of the normalized deviation $\sigma$ in the range $-1 < \sigma < 1$ refer to regions of primary eddy growth where the fractional volume dilution $k$ (Eq. 2) by eddy mixing process has to be taken into account for determining the probability distribution $P$ of fractal fluctuations (see Section 5.4 below).

## 5.4 Primary eddy growth region fractal space-time fluctuation probability distribution

Normalized deviation $\sigma$ ranging from $-1$ to $+1$ corresponds to the primary eddy growth region. In this region the probability $P$ is shown to be equal to $P = \tau^{-4k}$ (see below) where $k$ is the fractional volume dilution by eddy mixing (Eq. 2).

For the primary eddy growth region, the normalized deviation $\sigma$ represents the length step growth number for growth stages more than one. The first stage of eddy growth is the primary eddy growth starting from unit length scale perturbation, the complete eddy forming at the tenth length scale growth, i.e., $R = 10r$ and scale ratio $z$ equals 10. The steady state fractional volume dilution $k$ of the growing primary eddy by internal smaller scale eddy mixing is given by Eq. 2 as

$$k = \frac{w_* r}{WR} \tag{17}$$

The expression for $k$ in terms of the length scale ratio $z$ equal to $R/r$ is obtained from Eq. 1 as

$$k = \sqrt{\frac{\pi}{2z}} \qquad (18)$$

A fully formed large eddy length $R = 10r$ ($z$=10) represents the average or mean level zero and corresponds to a maximum of 50% probability of occurrence of either positive or negative fluctuation peak at normalized deviation σ value equal to zero by convention. For intermediate eddy growth stages, i.e., $z$ less than 10, the probability of occurrence of the primary eddy fluctuation does not follow conventional statistics, but is computed as follows taking into consideration the fractional volume dilution of the primary eddy by internal turbulent eddy fluctuations. Starting from unit length scale fluctuation, the large eddy formation is completed after 10 unit length step growths, i.e., a total of 11 length steps including the initial unit perturbation. At the second step ($z$ = 2) of eddy growth the value of normalized deviation σ is equal to 1.1 - 0.2 (= 0.9) since the complete primary eddy length plus the first length step is equal to 1.1. The probability of occurrence of the primary eddy perturbation at this σ value however, is determined by the fractional volume dilution $k$ which quantifies the departure of the primary eddy from its undiluted average condition and therefore represents the normalized deviation σ. Therefore the probability density $P$ of fractal fluctuations of the primary eddy is given using the computed value of $k$ as shown in the following equation.

$$P = \tau^{-4k} \qquad (19)$$

The vertical velocity $W$ spectrum will therefore be represented by the probability density distribution $P$ for values of $-1 \leq \sigma \leq 1$ given in Eq. 19 since fractal fluctuations exhibit quantum-like chaos as explained above (Eq. 16).

$$W = P = \tau^{-4k} \qquad (20)$$

The probabilities of occurrence ($P$) of the primary eddy for a complete eddy cycle either in the positive or negative direction starting from the peak value (σ = 0) are given for progressive growth stages (σ values) in the following Table 2. The statistical normal probability density distribution corresponding to the normalized deviation σ values are also given in the Table 2.

The model predicted probability density distribution $P$ along with the corresponding statistical normal distribution with probability values plotted on linear and logarithmic scales respectively on the left and right hand sides are shown in Figure 4. The model predicted probability distribution $P$ for fractal space-time fluctuations is very close to the statistical normal distribution for normalized deviation σ values less than 2 as seen on the left hand side of Figure 4. The model predicts progressively higher values of probability $P$ for values of σ greater than 2 as seen on a logarithmic plot on the right hand side of Figure 4.

| Table 2: Primary eddy growth | | | | |
|---|---|---|---|---|
| Growth step no | ± σ | k | Probability (%) | |
| | | | Model predicted | Statistical normal |
| 2 | .9000 | .8864 | 18.1555 | 18.4060 |
| 3 | .8000 | .7237 | 24.8304 | 21.1855 |
| 4 | .7000 | .6268 | 29.9254 | 24.1964 |
| 5 | .6000 | .5606 | 33.9904 | 27.4253 |
| 6 | .5000 | .5118 | 37.3412 | 30.8538 |
| 7 | .4000 | .4738 | 40.1720 | 34.4578 |
| 8 | .3000 | .4432 | 42.6093 | 38.2089 |
| 9 | .2000 | .4179 | 44.7397 | 42.0740 |
| 10 | .1000 | .3964 | 46.6250 | 46.0172 |
| 11 | 0 | .3780 | 48.3104 | 50.0000 |

fractal fluctuations probability distribution
comparison with statistical normal distribution

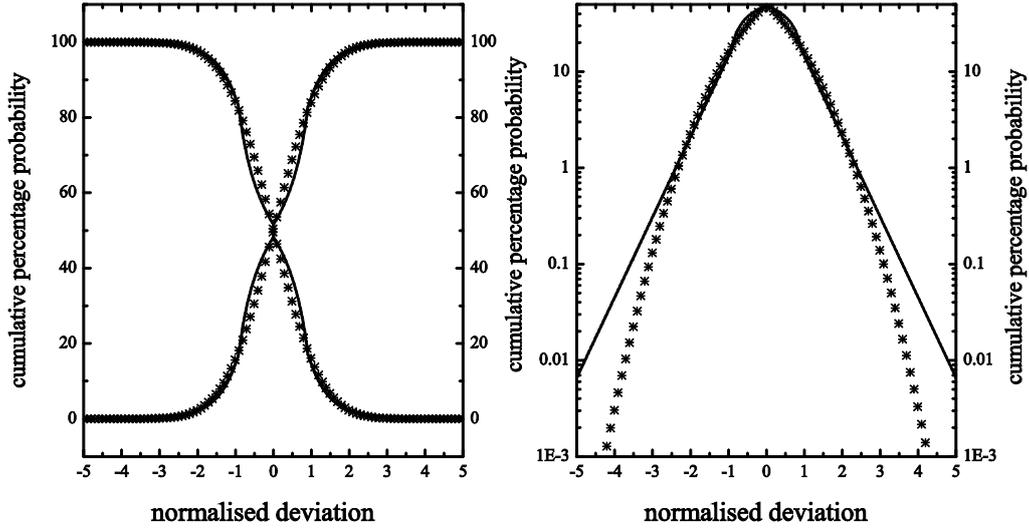

✳ statistical normal distribution ——— model predicted distribution

Fig. 4: Model predicted probability distribution $P$ along with the corresponding statistical normal distribution with probability values plotted on linear and logarithmic scales respectively on the left and right hand sides.

## 5.5 Atmospheric wind spectrum and aerosol size spectrum

The steady state flux $dN$ of cloud condensation nuclei (CCN) at level $z$ in the normalized vertical velocity perturbation $(dW)z$ is given as

$$dN = N(dW)z \qquad (21)$$

The logarithmic wind profile relationship for $W$ at Eq. 3 gives

$$\mathrm{d}N = Nz\frac{w_*}{k}\mathrm{d}(\ln z) \tag{22}$$

The general systems theory predicts universal logarithmic wind profile (Selvam 1990, Selvam and Fadnavis 1998) as manifested in the spiralling vortex air flows of tornadoes and the hurricane spiral cloud circulations.

Substituting for $k$ from Eq. 2

$$\mathrm{d}N = Nz\frac{w_*}{w_*}Wz\mathrm{d}(\ln z) = NWz^2\mathrm{d}(\ln z) \tag{23}$$

The length scale $z$ is related to the aerosol radius $r_a$ (Eq. 14). Therefore

$$\ln z = \frac{3}{2}\ln\left(\frac{r_a}{r_{as}}\right) \tag{24}$$

Defining a normalized radius $r_{an}$ equal to $\frac{r_a}{r_{as}}$, i.e., $r_{an}$ represents the CCN mean volume radius $r_a$ in terms of the CCN mean volume radius $r_{as}$ at the surface (or reference level). Therefore

$$\ln z = \frac{3}{2}\ln r_{an} \tag{25}$$

$$\mathrm{d}\ln z = \frac{3}{2}\mathrm{d}\ln r_{an} \tag{26}$$

Substituting for $\mathrm{d}\ln z$ in Eq. 23

$$\mathrm{d}N = NWz^2\frac{3}{2}\mathrm{d}(\ln r_{an}) \tag{27}$$

$$\frac{\mathrm{d}N}{\mathrm{d}(\ln r_{an})} = \frac{3}{2}NWz^2 \tag{28}$$

Substituting for $W$ from Eq. 16 and Eq. 20 in terms of the universal probability density $P$ for fractal fluctuations

$$\frac{\mathrm{d}N}{\mathrm{d}(\ln r_{an})} = \frac{3}{2}NPz^2 \tag{29}$$

The general systems theory predicts that fractal fluctuations may be resolved into an overall logarithmic spiral trajectory with the quasiperiodic Penrose tiling pattern for the internal structure such that the successive eddy lengths follow the Fibonacci mathematical series (Selvam 1990, 2007). The eddy length scale ratio $z$ for length step $\sigma$ is therefore a function of the golden mean $\tau$ given as

$$z = \tau^\sigma \tag{30}$$

Expressing the scale length $z$ in terms of the golden mean $\tau$ in Eq. 29

$$\frac{\mathrm{d}N}{\mathrm{d}(\ln r_{\mathrm{an}})} = \frac{3}{2} NP\tau^{2\sigma} \tag{31}$$

In Eq. 31 $N$ is the steady state aerosol concentration at level $z$. The normalized aerosol concentration at any level $z$ is given as

$$\frac{1}{N}\frac{\mathrm{d}N}{\mathrm{d}(\ln r_{\mathrm{an}})} = \frac{3}{2} P\tau^{2\sigma} \tag{32}$$

The fractal fluctuations probability density is $P = \tau^{-4\sigma}$ (Eq. 16) for values of the normalized deviation $\sigma \geq 1$ and $\sigma \leq -1$ on either side of $\sigma = 0$ as explained earlier (Section 5.3 and Section 5.4). Values of the normalized deviation $-1 \leq \sigma \leq 1$ refer to regions of primary eddy growth where the fractional volume dilution $k$ (Eq. 2) by eddy mixing process has to be taken into account for determining the probability density $P$ of fractal fluctuations. Therefore the probability density $P$ in the primary eddy growth region ($\sigma \geq 1$ and $\sigma \leq -1$) is given using the computed value of $k$ as $P = \tau^{-4k}$ (Eq. 20).

The normalized radius $r_{\mathrm{an}}$ is given in terms of $\sigma$ and the golden mean $\tau$ from Eq. 25 and Eq. 30 as follows.

$$\ln z = \frac{3}{2} \ln r_{\mathrm{an}}$$
$$r_{\mathrm{an}} = z^{2/3} = \tau^{2\sigma/3} \tag{33}$$

The normalized aerosol size spectrum is obtained by plotting a graph of normalized aerosol concentration $\frac{1}{N}\frac{\mathrm{d}N}{\mathrm{d}(\ln r_{\mathrm{an}})} = \frac{3}{2} P\tau^{2\sigma}$ (Eq. 32) versus the normalized aerosol radius $r_{\mathrm{an}} = \tau^{2\sigma/3}$ (Eq. 33). The normalized aerosol size spectrum is derived directly from the universal probability density $P$ distribution characteristics of fractal fluctuations (Eq. 16 and Eq. 20) and is independent of the height $z$ of measurement and is universal for aerosols in turbulent atmospheric flows. The aerosol size spectrum is computed starting from the minimum size, the corresponding probability density $P$ (Eq. 32) refers to the cumulative probability density starting from 1 and is computed as equal to $P = 1 - \tau^{-4\sigma}$. The universal normalized aerosol size spectrum represented by $\frac{1}{N}\frac{\mathrm{d}N}{\mathrm{d}(\ln r_{\mathrm{an}})}$ versus the normalized aerosol radius $r_{\mathrm{an}}$ is shown in Figure 5.

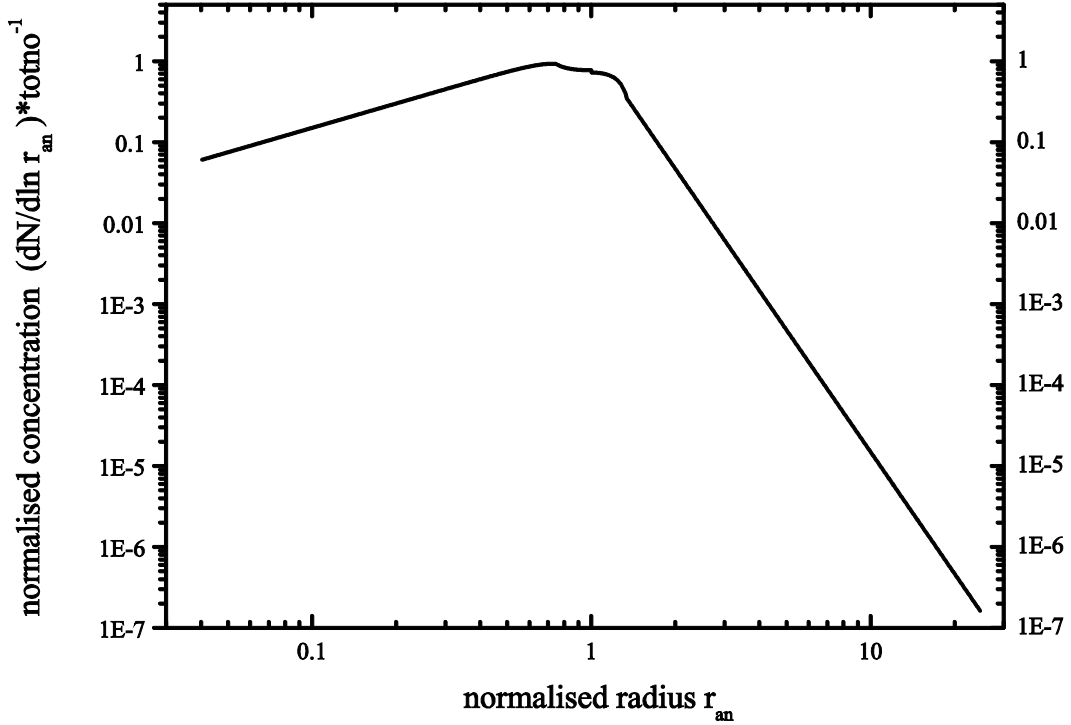

Fig. 5: Model predicted universal (scale independent) aerosol size spectrum

## 6. General Systems Theory and Maximum Entropy Principle of Classical Statistical Physics

Kaniadakis (2009) states that the correctness of an analytic expression for a given power-law tailed distribution used to describe a statistical system, is strongly related to the validity of the generating mechanism. In this sense the maximum entropy principle, the cornerstone of statistical physics, is a valid and powerful tool to explore new roots in searching for generalized statistical theories (Kaniadakis, 2009). The concept of entropy is fundamental in the foundation of statistical physics. It first appeared in thermodynamics through the second law of thermodynamics. In statistical mechanics, we are interested in the disorder in the distribution of the system over the permissible microstates. The measure of disorder first provided by Boltzmann principle (known as Boltzmann entropy) is given by $S = K_B \ln M$, where $K_B$ is the thermodynamic unit of measurement of entropy and is known as Boltzmann constant equal to $1.33 \times 10^{-16}$ erg/$^\circ$C. The variable $M$, called thermodynamic probability or statistical weight, is the total number of microscopic complexions compatible with the macroscopic state of the system and corresponds to the "degree of disorder" or 'missing information' (Chakrabarti and De, 2000).

The maximum entropy principle concept of classical statistical physics is applied to determine the fidelity of the inverse power law probability distribution $P$ (Eq. 15) for exact quantification of the observed space-time fractal fluctuations of dynamical systems ranging from the microscopic dynamics of quantum systems to macro-scale real world systems. The eddy energy probability distribution ($P$) of fractal space-time fluctuations for each stage of hierarchical eddy growth is given by Eq. (15) derived earlier, namely

$$P = \tau^{-4\,t}$$

The r.m.s circulation speed $W$ of the large eddy follows a logarithmic relationship with respect to the length scale ratio $z$ equal to $R/r$ (Eq. 3) as given below

$$W = \frac{w_*}{k} \log z$$

In the above equation the variable $k$ represents for each step of eddy growth, the fractional volume dilution of large eddy by turbulent eddy fluctuations carried on the large eddy envelope (Selvam, 1990) and is given as (Eq. 2)

$$k = \frac{w_* r}{WR}$$

Substituting for $k$ in Eq. (3) we have

$$W = w_* \frac{WR}{w_* r} \log z = \frac{WR}{r} \log z$$

*and* $\qquad\qquad\qquad\qquad\qquad\qquad\qquad\qquad$ (34)

$$\frac{r}{R} = \log z$$

The ratio $r/R$ represents the fractional probability $P$ of occurrence of small-scale fluctuations ($r$) in the large eddy ($R$) environment. Since the scale ratio $z$ is equal to $R/r$, Eq. (34) may be written in terms of the probability $P$ as follows.

$$\frac{r}{R} = \log z = \log\left(\frac{R}{r}\right) = \log\left(\frac{1}{(r/R)}\right)$$

$$P = \log\left(\frac{1}{P}\right) = -\log P$$

$\qquad\qquad\qquad\qquad\qquad\qquad\qquad\qquad$ (35)

For a probability distribution among a discrete set of states the generalized entropy for a system out of equilibrium is given as (Salingaros and West, 1999; Chakrabarti and De, 2000; Beck, 2009; Sethna, 2009)

$$S = -\sum_{j=1}^{n} P_j \ln P_j$$
$\qquad\qquad\qquad\qquad\qquad\qquad\qquad\qquad$ (36)

In Eq. (36) $P_j$ is the probability for the $j^{\text{th}}$ stage of eddy growth in the present study and the entropy $S$ represents the 'missing information' regarding the probabilities. Maximum entropy $S$ signifies minimum preferred states associated with scale-free probabilities.

The validity of the probability distribution $P$ (Eq. 15) is now checked by applying the concept of maximum entropy principle (Kaniadakis, 2009). Substituting for $\log P_j$ (Eq. 36) and for the probability $P_j$ in terms of the golden mean $\tau$ derived earlier (Eq. 15) the entropy $S$ is expressed as

$$S = -\sum_{j=1}^{n} P_j \log P_j = \sum_{j=1}^{n} P_j^2 = \sum_{j=1}^{n} \left(\tau^{-4n}\right)^2$$

$$S = \sum_{j=1}^{n} \tau^{-8n} \approx 1 \text{ for large } n \tag{37}$$

In Eq. (37) $S$ is equal to the square of the cumulative probability density distribution and it increases with increase in $n$, i.e., the progressive growth of the eddy continuum and approaches 1 for large $n$. According to the second law of thermodynamics, increase in entropy signifies approach of dynamic equilibrium conditions with scale-free characteristic of fractal fluctuations and hence the probability distribution $P$ (Eq. 15) is the correct analytic expression quantifying the eddy growth processes visualized in the general systems theory. The ordered growth of the atmospheric eddy continuum is associated with maximum entropy production.

Paltridge (2009) states that the principle of maximum entropy production (MEP) is the subject of considerable academic study, but is yet to become remarkable for its practical applications. The ability of a system to dissipate energy and to produce entropy "ought to be" some increasing function of the system's structural complexity. It would be nice if there were some general rule to the effect that, in any given complex system, the steady state which produces entropy at the maximum rate would at the same time be the steady state of maximum order and minimum entropy (Paltridge, 2009).

Selvam (2011) has shown that the eddy continuum energy distribution $P$ (Eq. 15) is the same as the *Boltzmann distribution* for molecular energies. The derivation of *Boltzmann's equation* from general systems theory concepts visualises the eddy energy distribution as follows: (1) The primary small-scale eddy represents the molecules whose eddy kinetic energy is equal to $K_B T$ where $K_B$ is the Boltzmann's constant and $T$ the temperature as in classical physics. (2) The energy pumping from the primary small-scale eddy generates growth of progressive larger eddies (Selvam, 1990). The r.m.s circulation speeds $W$ of larger eddies are smaller than that of the primary small-scale eddy (Eq. 1). (3) The space-time *fractal* fluctuations of molecules (atoms) in an ideal gas may be visualized to result from an eddy continuum with the eddy energy $E$ per unit volume relative to primary molecular kinetic energy $K_B T$ decreasing progressively with increase in eddy size.

The eddy energy probability distribution ($P$) of fractal space-time fluctuations also represents the *Boltzmann distribution* for each stage of hierarchical eddy growth and is given by Eq. (15) derived earlier, namely

$$P = \tau^{-4t}$$

The general systems theory concepts are applicable to all space-time scales ranging from microscopic scale quantum systems to macroscale real world systems such as atmospheric flows.

A systems theory approach based on maximum entropy principle has been applied in cloud physics to obtain useful information on droplet size distributions without regard to the details of individual droplets (Liu et al. 1995; Liu 1995; Liu and Hallett 1997, 1998; Liu and Daum, 2001; Liu, Daum and Hallett, 2002; Liu, Daum, Chai and Liu, 2002). Liu, Daum et al. (2002) conclude that a combination of the systems idea with multiscale approaches seems to be a promising avenue. Checa and Tapiador (2011) have presented a maximum entropy approach to Rain Drop Size Distribution (RDSD) modelling. Liu, Liu and Wang (2011) have given a review of the concept of entropy and its relevant principles, on the organization of atmospheric systems and the principle of the Second Law of thermodynamics, as well as their applications to atmospheric sciences. The Maximum Entropy Production Principle (MEPP), at least as used in climate science, was first hypothesised by Paltridge (1975, 1979).

# 7. Data

Four data sets, namely, two aerosol (I and II), one cloud drop size (III) and one rain drop size (IV) were used for comparison of observed with model predicted suspended particle size spectrum in turbulent atmospheric flows.

## 7.1 Data set I, aerosol size spectrum

TARFOX_WALLOPS_SMPS: Tropospheric Aerosol Radiative Forcing Observational eXperiment (TARFOX), Langley DAAC Project - Scanning Mobility Particle Sizer (TSI Incorporated, St. Paul, MN) data taken at Wallops ground station ($37.85^o$ lat, $-75.48^o$ lon) in the U.S. eastern seaboard. Ground-based ambient size distribution of aerosol (10.7 to 749 nm diameter, fine mode) at point measurements at 5 minutes time intervals was taken during the period $10^{th}$ to $31^{st}$ July 1996. Raw data exported using SMPS 2.0 (TSI), then imported to Microsoft Excel, adjusted to local time, then saved as comma delimited. Data were obtained from http://eosweb.larc.nasa.gov/PRODOCS/tarfox/table_tarfox.html and http://eosweb.larc.nasa.gov/cgi-bin/searchTool.cgi?Dataset=TARFOX_WALLOPS_SMPS

## 7.2 Data Set II, aerosol size spectrum

PCASP files (replaced on January 14, 2005): Contain 1 Hz size distribution data measured aboard the CIRPAS (Center for Interdisciplinary Remotely-Piloted Aircraft Studies) Twin Otter during the Atmospheric Radiation Program (ARM) Intensive Operational Period (IOP) 2003 using the PCASP, with SPP-200 electronics. PCASP is Passive Cavity Aerosol Spectrometer manufactured by PMS Inc., but with a SPP-200 data system manufactured by DMT Inc. Ponca City, Oklahoma, USA was the base from which the flights were conducted. These were typically between 3 and 5 hour flights, carried out during the month of May 2003. The Aerosol IOP was conducted between May 5-31, 2003 over the ARM Southern Great Plains (SGP) site. There were a total of 16 science flights, for a total of 60.6 flight hours, conducted by the CIRPAS Twin Otter aircraft on 15 days during this period. Most of the Twin Otter flights were conducted under clear or partly cloudy skies to assess aerosol impacts on solar radiation.

 The aerosol particle size concentrations ranging in diameter from 0.1µm to 3.169701µm (accumulation and coarse modes) were measured in 20 channels (size ranges). The geometric mean radius of the class interval was used for computing d(ln$r_{an}$). The data sets were obtained from ARM IOP Data Archive http://www.archive.arm.gov/armlogin/login.jsp.

## 7.3 Data Set III, cloud drop size/number concentration

Cloud drop size/number concentration. Project SAFARI 2000, CARG Aerosol and Cloud Data from the Convair-580. Web Site: http://cargsun2.atmos.washington.edu/. The Cloud and Aerosol Research Group (CARG) of the University of Washington participated in the SAFARI-2000 Dry Season Aircraft campaign with their Convair-580 research aircraft. This campaign covered five countries in southern Africa from 10 August through 18 September, 2000 on the thirty-one research flights. (http://daac.ornl.gov/data/safari2k/atmospheric/CV-580/comp/SAFARI-MASTER.pdf).

Data Citation: Hobbs, P. V. 2004. SAFARI 2000 CV-580 Aerosol and Cloud Data, Dry Season 2000 (CARG). Data set. Available on-line (http://www.daac.ornl.gov) from Oak Ridge National Laboratory Distributed Active Archive Center, Oak Ridge, Tennessee, U.S.A. doi:10.3334/ORNLDAAC/710. All Data Taken At Latitude: 14.00S To 26.00S, Longitude: 36.00E To 11.00E.

Data details: Cloud particle concentration per cc between 1.7 and 47.0 μm in 15 channels. Particle Measuring Systems Model FSSP-100. Calculated from raw counts and sample time. Seven data sets containing cloud drop size/number concentration were used for the study. The data are output at 1-second resolution (http://daac.ornl.gov/S2K/guides/s2k_CV580.html).

### 7.4 Data Set IV: TWP-ICE, Joss-Waldvogel Disdrometer raindrop size distributions.

The Tropical Western Pacific – International Cloud Experiment (TWP-ICE) was held near Darwin, Australia to collect in-situ and remote-sensing measurements of clouds, precipitation, and meteorological variables from the ground to the lower stratosphere. During TWP-ICE, vertically-pointing profiling radar, surface rain gauge, and disdrometer observations were collected for the whole wet season from November 2005 through March 2006. The Joss-Waldvogel Disdrometer was operational from 3 November 2005 through 10 February 2006.

## 8. Analysis and discussion of results

The atmospheric suspended particulate size spectrum is closely related to the vertical velocity spectrum (Section 5). The mean volume radius of suspended aerosol particulates increases with height (or reference level $z$) in association with decrease in number concentration. At any height (or reference level) $z$, the fractal fluctuations (of wind, temperature, etc.) carry the signatures of eddy fluctuations of all size scales since the eddy of length scale $z$ encloses smaller scale eddies and at the same time forms part of internal circulations of eddies larger than length scale $z$ (Section 5.2). The observed atmospheric suspended particulate size spectrum also exhibits a decrease in number concentration with increase in particulate radius. At any reference level $z$ of measurement the mean volume radius $r_{as}$ will serve to calculate the normalized radius $r_{an}$ for the different radius class intervals as explained below.

The general systems theory for fractal space-time fluctuations in dynamical systems predicts universal mass size spectrum for atmospheric suspended particulates (Section 5). For homogeneous atmospheric suspended particulates, i.e. with the same particulate substance density, the atmospheric suspended particulate mass and radius size spectrum is the same and is given as (Section 5.5) the normalized aerosol number concentration equal to $\dfrac{1}{N}\dfrac{\mathrm{d}N}{\mathrm{d}(\ln r_{an})}$ versus the normalized aerosol radius $r_{an}$, where (i) $r_{an}$ is equal to $\dfrac{r_a}{r_{as}}$, $r_a$ being the mean class interval radius and $r_{as}$ the mean volume radius for the total aerosol size spectrum (ii) $N$ is the total aerosol number concentration and d$N$ is the aerosol number concentration in the aerosol radius class interval $dr_a$ (iii) d(ln $r_{an}$) is equal to $\dfrac{dr_a}{r_a}$ for the aerosol radius class interval $r_a$ to $r_a + dr_a$.

## 8.1 Analysis results, Data I: TARFOX_WALLOPS_SMPS, aerosol size spectra

A total of 23 data sets between 16 July and 26 July 1996 are available for the study. The data consists of particle number concentration per cc in 59 class intervals ranging from 10.7 to 749 nm (fine mode) for the particle diameter. The mid-point diameter of the class interval was used to compute the corresponding value of d(ln$r_{an}$). The average aerosol size spectra for each of the 23 data sets are plotted on the left hand side and the total average spectrum for the 23 data sets is plotted on the right and side in Figure 6a along with the model predicted scale independent aerosol size spectrum. The corresponding standard deviations for the average spectra are shown as error bars in Figs. 6a. The average values of mean volume radius (nm), total number concentration (cm$^{-3}$), the number of spectra, the number of particle size class intervals for each of the 23 data sets and the upper and lower bounds of particle size (diameter in nm) intervals are given in Figure 6b.

The aerosol size spectra cover the fine size range and displays large deviations from the mean, particularly for the individual flights (left hand side of Fig. 6a). The source of the uncertainties displayed by the error bars (Fig. 6a) may be due to measurement noise, independent in every size interval, also may be due to different aerosol sources with different particle substance densities. However, a major portion of the total average aerosol size spectrum (right hand side of Fig. 6a) shows a reasonably good fit (within plus or minus two standard deviations) to the model predicted universal spectrum for homogeneous aerosols, i.e. same aerosol source.

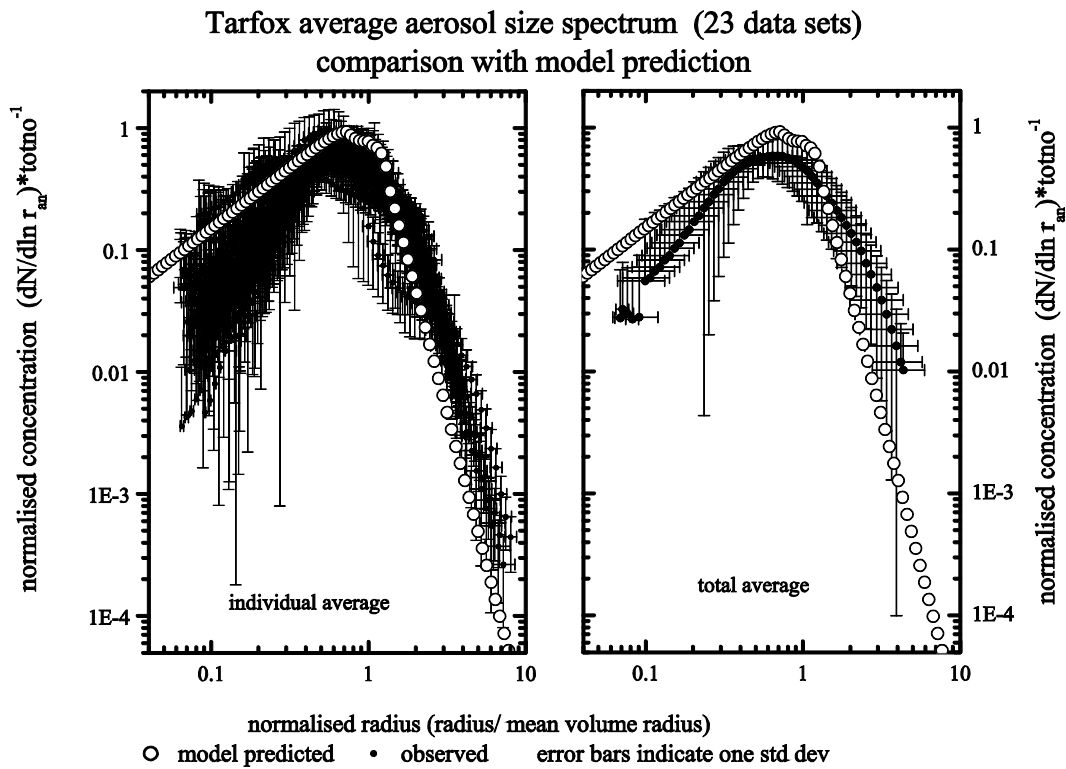

Tarfox average aerosol size spectrum (23 data sets) comparison with model prediction

Fig. 6a Average aerosol size spectrum for each of the 23 data sets (left) and total average aerosol size spectrum for the 23 data sets (right).. Error bars indicate one standard deviation on either side of the mean. Model predicted scale independent aerosol size spectrum also is shown in the figure.

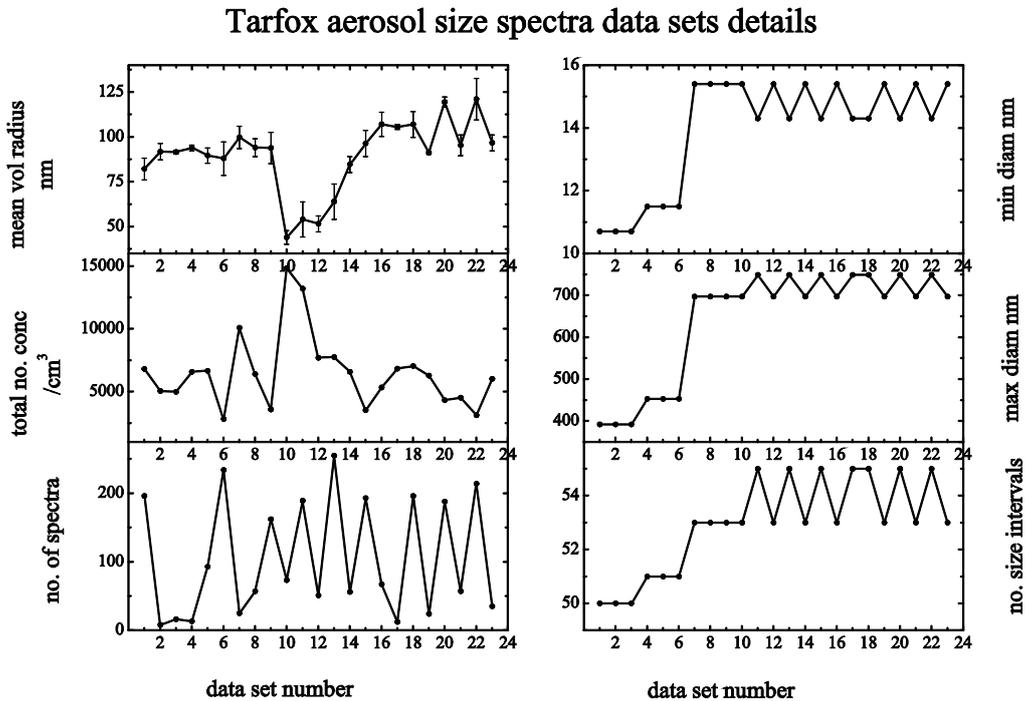

Fig. 6b Details of (i) mean volume radius nm (ii) aerosol total number concentration cm$^{-3}$ (iii) number of spectra (iv) lower bound (v)upper bound diameter nm for the particle size class intervals used (total available=59) (vi) number of particle size class intervals for each of the 23 data sets available for TARFOX aerosol size spectra

## 8.2 Analysis results, Data II: CIRPAS Twin Otter flight data sets, aerosol size spectra

ARM Aerosol IOP at the Southern Great Plains (SGP) site over a 3-4 week period centered on May 2003. CIRPAS Twin Otter flight data sets using the PCASP. A total of 16 data sets are available for the study.

The aerosol particle size concentrations ranging in diameter from 0.1μm to 3.169701μm consisting of accumulation mode (upto 1μm) and coarse mode (diameter > 1μm) were measured in 20 channels (size ranges). The geometric mean radius of the class interval was used for computing d(ln$r_{an}$). The data sets were obtained from ARM IOP Data Archive http://www.archive.arm.gov/armlogin/login.jsp. The average aerosol size spectra for each of the 16 data sets are plotted on the left hand side and the total average spectrum for the 16 data sets is plotted on the right and side in Figure 7a along with the model predicted scale independent aerosol size spectrum. The corresponding standard deviations for the average spectra are shown as error bars in Figs. 7a. The average values of mean volume radius, total number concentration, the number of spectra and the upper and lower bounds of particle size (radius) intervals are given in Figure 7b.

The portion of the total average aerosol size spectrum corresponding to the accumulation mode (upto 1μm or normalized diameter equal to 4) shows a reasonably good fit (within plus or minus two standard deviations) to the model predicted universal spectrum. The source of the uncertainties displayed by the error bars may be due to measurement noise, independent in every size interval, also may be due to different aerosol sources. The coarse mode (diameter > 1μm) portion of the total average aerosol size spectrum shows significant departure from the model predicted spectrum and may be attributed to a different source region for the suspended particulates with a different density. The model predicts universal spectrum for suspended aerosol mass size distribution (Section 5), based on the concept that the atmospheric eddies hold in suspension the aerosols and thus the mass size spectrum of the atmospheric aerosols is dependent on the vertical velocity fluctuation spectrum of the atmospheric eddies.

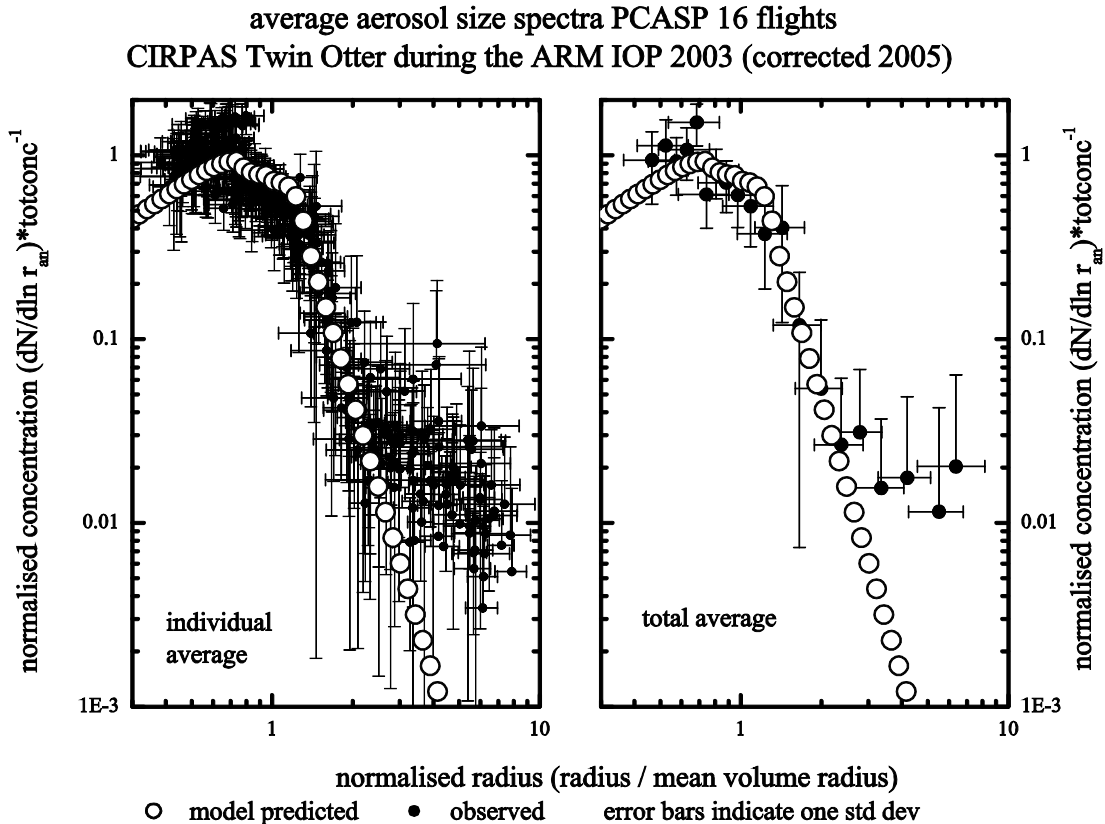

Fig. 7a Average aerosol size spectrum for each of the 16 data sets (left) and total average aerosol size spectrum for the 16 data sets (right).. Error bars indicate one standard deviation on either side of the mean. Model predicted scale independent aerosol size spectrum also is shown in the figure.

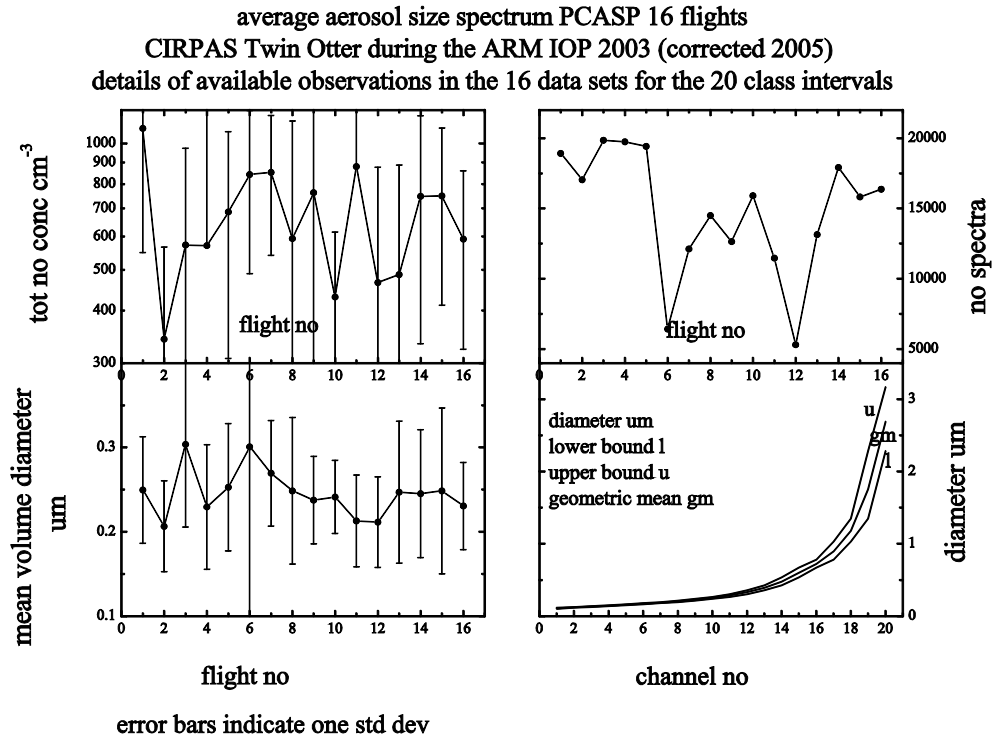

Fig. 7b Details of (i) mean volume diameter µm (ii) aerosol total number concentration cm$^{-3}$ and (iii) number of spectra for the 16 data sets and the lower bound, upper bound and geometric mean diameter µm for the 20 particle size class intervals.

## 8.3 Analysis results, Data III: CARG aerosol and cloud data from the Convair-580, cloud drop size spectra

Cloud drop size/number concentration from Project SAFARI 2000, CARG Aerosol and Cloud Data from the Convair-580. A total of 7 data sets are available for the study. Cloud drop size spectra were computed for data sets for which the cloud liquid water content (Johnson-Williams) was more than zero.

The cloud drop size/number concentrations ranging in diameter from 1.70µm to 47.0µm were measured in 15 channels (size ranges). The arithmetic mean radius of the class interval was used for computing d(ln$r_{an}$). The data sets were obtained from ftp://ftp.daac.ornl.gov/data/safari2k/atmospheric/CV-580/data/. The average aerosol size spectra for each of the 7 data sets are plotted on the left hand side and the total average spectrum for the 7 data sets is plotted on the right and side in Figure 8a along with the model predicted scale independent aerosol size spectrum. The corresponding standard deviations for the average spectra are shown as error bars in Figs. 8a. The average values of mean volume radius, total number concentration, the number of spectra, the number of class intervals for each of the 7 data sets and the upper and lower bounds of particle size (radius) intervals are given in Figure 8b.

The individual and total average cloud droplet spectra show a close fit to model predicted universal spectrum for cloud drop diameter more than 5 µm (corresponding to normalised diameter equal to 1). The observed spectra show appreciably larger radii than model predicted for normalized radius size range less than 1 and may be attributed to the large increase in the sampled median volume diameter, from about 5 µm to 20 µm during flights 5 to 7 (Fig. 8b). However, even in this region for normalized radius less than 1, the model predicted and the total average spectrum (right hand side of Fig. 8a) are within two standard deviations from the mean, a standard statistical criterion for 'goodness of fit'.

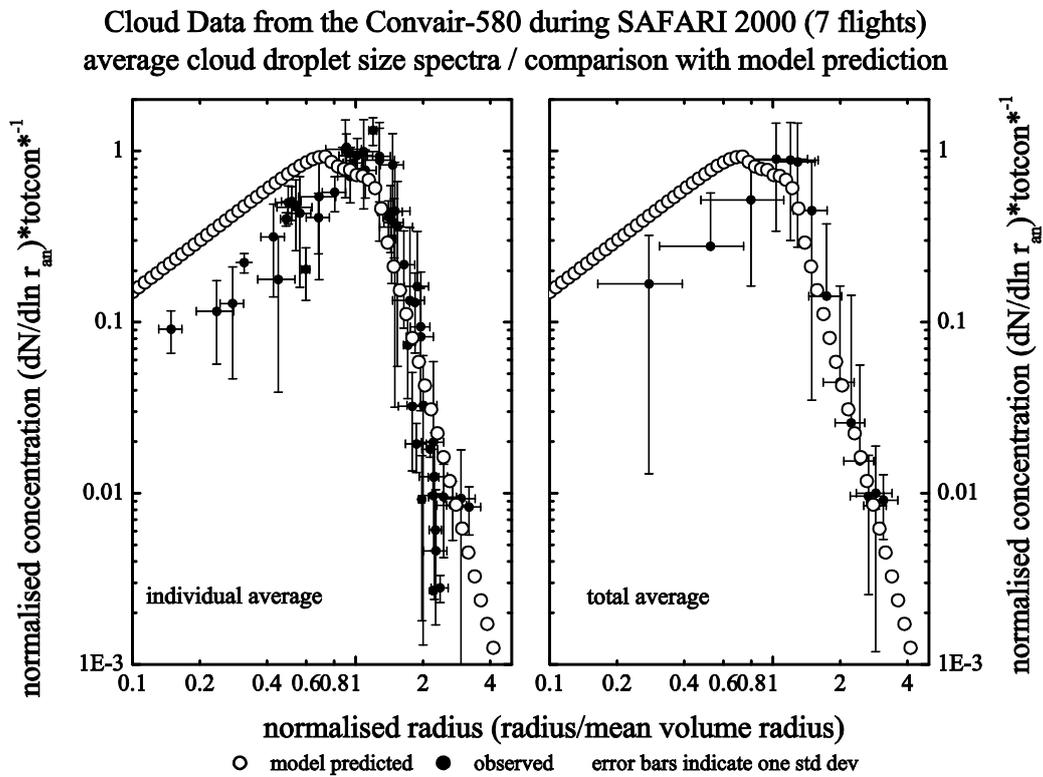

Fig. 8a Average cloud drop size spectrum for each of the 7 data sets (left) and total average cloud drop size spectrum for the 7 data sets (right).. Error bars indicate one standard deviation on either side of the mean. Model predicted scale independent suspended particulate size spectrum also is shown in the figure.

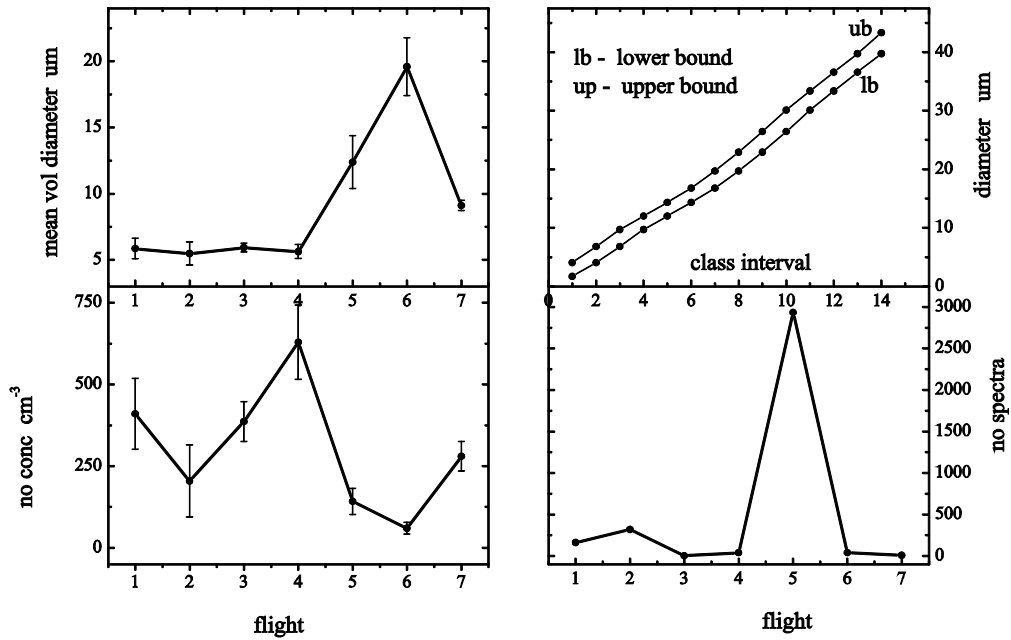

Fig. 8b Details of (i) mean volume diameter um (ii) cloud drop total number concentration cm$^{-3}$ and (iii) number of spectra for each of the 7 data sets and the lower bound and upper bound diameter μm for the 15 drop size class intervals.

## 8.4 Analysis results Data IV: TWP-ICE, Joss-Waldvogel Disdrometer raindrop size distributions.

The Joss-Waldvogel Disdrometer (http://cires.colorado.edu/blogs/twc-ice/2009/09/19/4-0-joss-waldvogel-disdrometer) was operational from 3 November 2005 through 10 February 2006. The original JWD data were collected at the full 127 diameter channels and with a 10 second dwell time. These high resolution data were reduced to the standard 20 diameter channels and to a 1-minute resolution. A dead-time correction (Sheppard and Joe 1994; Sauvageot and Lacaux 1995) was applied to the raindrop counts. The data products are provided using the dead-time corrected raindrop counts and 60 second dwell time.

The ASCII data files are day files with 1-minute resolution and contain 1440 rows. Bad or missing data values are indicated with a value of -99.9. The ASCII data files can be found on the ftp site: ftp://ftp.etl.noaa.gov/user/cwilliams/Darwin/disdrometer/dat/

The values of d(lnr$_{an}$) for the rain drop size spectrum was calculated from the number of raindrops in each raindrop diameter size in a total of 20 standard diameter channels ranging from 0.34mm to 5.37mm and corresponding channel width (mm). A total of 99 data sets (days) are available for the study. The average (daily) rain drop size spectra for each of the 99 data sets are plotted on the left hand side and the total average (daily) spectrum for the 99 data sets is plotted on the right hand side in Figure 9a along with the model predicted scale independent aerosol size spectrum. The corresponding standard deviations for the average spectra are shown as error bars in Figures. 9a. The average values of (i) mean volume diameter mm (ii) rain drop total number (iii) number of observations in each of the 20 channels and (iv) channel diameter and channel width (mm) for the 20 drop size channels are given in Figure 9b.

The total average raindrop size spectrum (right hand side of Fig. 9a) shows a reasonably good fit (within two standard deviations from the mean) even though the individual average raindrop size spectra (left hand side of Fig. 9a) show large error bars which may be attributed to the large variability in total number of drops sampled corresponding to each mean volume radius (Fig. 9b).

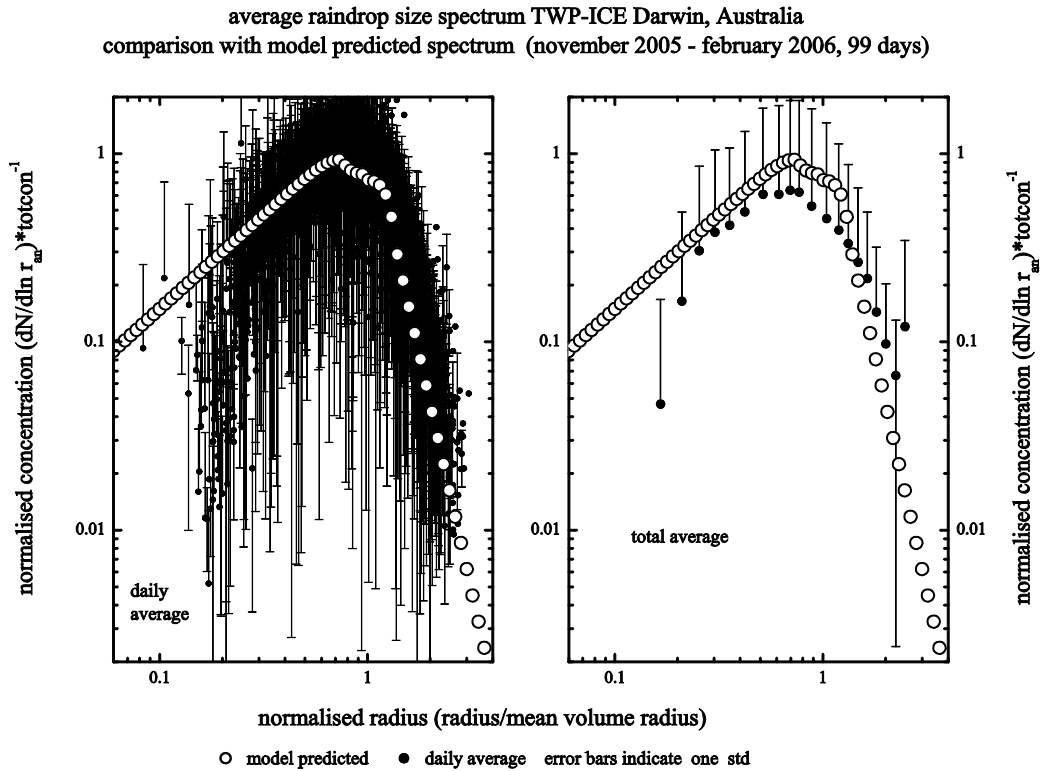

Fig. 9a The average (daily) rain drop size spectra for each of the 99 data sets are plotted on the left hand side and the total average (daily) spectrum for the 99 data sets is plotted on the right hand side in Fig. 9a along with the model predicted scale independent aerosol size spectrum. The corresponding standard deviations for the average spectra are shown as error bars in Figs. 9a.

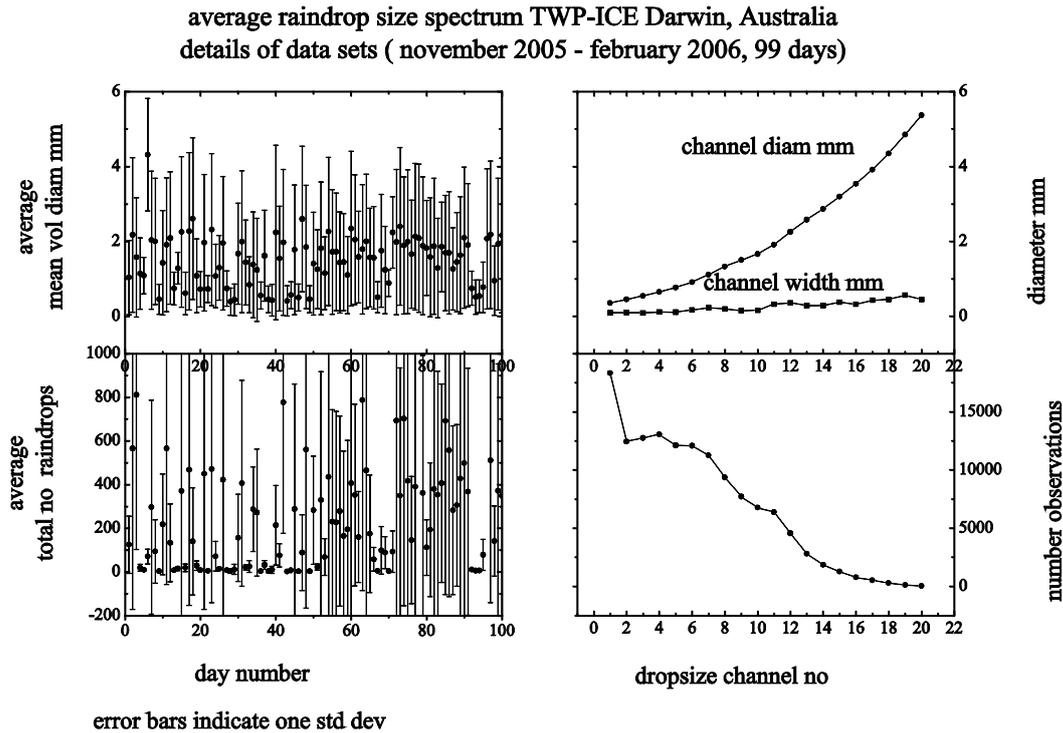

Fig. 9b The average values of (i) mean volume diameter mm (ii) rain drop total number (iii) number of observations in each of the 20 channels and (iv) channel diameter and channel width (mm) for the 20 drop size channels

## 9. Conclusions

A general systems theory for fractal space-time fluctuations in turbulent atmospheric flows predicts a universal scale-independent mass or radius size distribution for homogeneous suspended atmospheric particulates expressed as a function of the golden mean τ, the total number concentration and the mean volume radius. Model predicted spectrum is in agreement (within two standard deviations on either side of the mean) with total averaged radius size spectra for the following four experimentally determined data sets: (i) CIRPAS mission TARFOX_WALLOPS_SMPS aerosol size distributions (ii) CIRPAS mission ARM-IOP (Ponca City, OK) aerosol size distributions (iii) SAFARI 2000 CV-580 (CARG Aerosol and Cloud Data) cloud drop size distributions and (iv) TWP-ICE (Darwin, Australia) rain drop size distributions. SAFARI 2000 aerosol size distributions reported by Haywood et al. (2003) also show similar shape for the distributions. Classical statistical physical concepts underlie the physical hypothesis relating to the dynamics of the atmospheric eddy systems proposed in the present paper. Model predicted spectrum satisfies the maximum entropy principle of statistical physics.

The general systems theory model for aerosol size distribution is scale free and is derived directly from atmospheric eddy dynamical concepts. At present empirical models such as the log normal distribution with arbitrary constants for the size distribution of atmospheric suspended particulates are used for quantitative estimation of earth-atmosphere radiation budget related to climate warming/cooling trends (Section 2.1).

## Acknowledgement

The author is grateful to Dr. A. S. R. Murty for encouragement during the course of the study.

# Appendix I

## List of frequently used Symbols

$\nu$      frequency
$d$      aerosol diameter
$N$      aerosol number concentration
$N_*$      surface (or initial level) aerosol number concentration
$r_a$      aerosol radius
$\alpha$      exponent of inverse power law
$W$      circulation speed (root mean square) of large eddy
$w$      circulation speed (root mean square) of turbulent eddy
$R$      radius of the large eddy
$r$      radius of the turbulent eddy
$w_*$      primary (initial stage) turbulent eddy circulation speed
$r_*$      primary (initial stage) turbulent eddy radius
$T$      time period of large eddy circulation
$t$      time period of turbulent eddy circulation
$k$      fractional volume dilution rate of large eddy by turbulent eddy fluctuations
$z$      eddy length scale ratio equal to $R/r$
$f$      steady state fractional upward mass flux of surface (or initial level) air
$q$      moisture content at height $z$
$q_*$      moisture content at primary (initial stage) level
$m$      suspended aerosol mass concentration at any level $z$
$m_*$      suspended aerosol mass concentration at primary (initial stage) level
$r_a$      mean volume radius of aerosols at level $z$
$r_{as}$      mean volume radius of aerosols at primary (initial stage) level
$r_{an}$      normalized mean volume radius equal to $r_a/r_{as}$
$P$      probability density distribution of fractal fluctuations
$\sigma$      normalized deviation